\documentclass[galaxies,article,submit,moreauthors,pdftex]{Definitions/mdpi}

\firstpage{1}
\pubvolume{1}
\issuenum{1}
\articlenumber{0}
\pubyear{2026}
\copyrightyear{2026}
\datereceived{}
\daterevised{}
\dateaccepted{}
\datepublished{}

\usepackage[normalem]{ulem} %

\newcommand{\lat}{\textit{Fermi}-LAT}
\newcommand{\src}{FL16Y~J1908.6+0915e}
\newcommand{\magnetar}{SGR~1900+14}
\newcommand{\wf}{W49A}

\newcommand{\lhaaso}{LHAASO~J1907+0913}

\newcommand{\ergcms}{\mathrm{erg\,cm^{-2}\,s^{-1}}}
\newcommand{\gevcms}{\mathrm{GeV\,cm^{-2}\,s^{-1}}}
\newcommand{\Wp}{W_{\mathrm{p}}}

\Title{Two-Component Model for the Unidentified Extended \textit{Fermi}-LAT Source
4FGL J1908.6+0915e / FL16Y J1908.6+0915e
in the SGR~1900+14--W49A Sky Region}

\Author{Vadym Voitsekhovskyi $^{1}$*\orcidA{}, Roman Hnatyk $^{2}$\orcidB{}, Bohdan Hnatyk$^{2}$\orcidC{},
 Lidiia Zadorozhna $^{3,4}$\orcidD{}, Anastasiia Sokareva $^{5}$\orcidE{}, Valentyna Babur$^{6}$\orcidF{}, Olexandr Gugnin$^{5}$\orcidG{}}

\AuthorNames{Vadym Voitsekhovskyi, Roman Hnatyk, Bohdan Hnatyk, Lidiia Zadorozhna, Anastasiia Sokareva, Valentyna Babur and Olexandr Gugnin}

\address{%
$^{1}$ \quad Anton Pannekoek Institute / GRAPPA, University of Amsterdam, Science Park 904, 1098 XH Amsterdam, The Netherlands; v.voitsekhovskyi@uva.nl\\
$^{2}$ \quad Astronomical Observatory of Taras Shevchenko National University of Kyiv, Observatorna Str. 3,\linebreak 02000 Kyiv, Ukraine; roman.hnatyk@knu.ua (R.H.); bhnatyk@knu.ua (B.H.)\\
$^{3}$ \quad Niels Bohr Institute, University of Copenhagen, Jagtvej 155A, 2200 K{\o}benhavn, Denmark\\
$^{4}$ \quad Department of Quantum Field Theory and Astroparticle Physics, Faculty of Physics, Taras Shevchenko National University of Kyiv, Hlushkova Avenue 4, 03127 Kyiv, Ukraine; lidiia.zadorozhna@knu.ua \\
$^{5}$ \quad Department of Astronomy and Space Physics, Faculty of Physics, Taras Shevchenko National University of Kyiv, Hlushkova Avenue 4, 03127 Kyiv, Ukraine; sokareva.anastasiia@knu.ua (A.S.); alexgugnin@knu.ua (O.G.) \\
$^{6}$ \quad Deutsches Elektronen-Synchrotron (DESY), Platanenallee 6, 15738 Zeuthen, Germany; valentyna.babur@desy.de\\
}
\corres{Correspondence: v.voitsekhovskyi@uva.nl}

\abstract{Galactic PeVatrons are expected to be powerful sources of non-thermal
high-energy (HE, $E>100$~MeV) and very-high-energy (VHE, $E>100$~GeV) $\gamma$-ray
emission as a manifestation of acceleration processes in supernova remnants (SNRs),
star formation regions (SFRs), and pulsar wind nebulae (PWNe). In the sky region
containing the magnetar \magnetar{} and the young massive star cluster (YMSC) \wf{},
 there is one recently identified point-like source 
FL16Y~J1910.2+0903, and there are three unidentified gamma-ray sources: the \lat{} extended 
source \src{}, the HAWC extended source 4HWC~J1908+0845, and the
LHAASO point-like source \lhaaso{}. We investigate the possibility of explaining 
the HE and VHE $\gamma$-ray emission from the \magnetar{}--\wf{} sky region
 using a model of the \lat{} extended source \src{} as superposition of two extended sources 
- a potential SNR/PWN associated with the magnetar \magnetar{} and the stellar-wind 
superbubble adjacent to the YMSC \wf{} complex. To this end, we simulate the observed
$\gamma$-ray emission in hadronic and leptonic models
using the open NAIMA code and show that it consists of two extended regions: 
the stellar wind superbubble in \wf{} region -- an analog of the Westerlund~1 superbubble -- powered by the 
YMSC \wf{} (the southern part of the extended source \src{}); and the
\magnetar{} region, powered by a potential SNR/PWN associated with the magnetar
(the northern part of the extended source \src{}). However, 
due to the considerable distance to the
magnetar ($d=12.5$~kpc), the required cosmic-ray energy budget in both hadronic and
leptonic scenarios corresponds to a hypernova explosion and/or a powerful magnetar wind
nebula created by a newly born millisecond magnetar with a rotational energy of the
order of $10^{52}$~erg.}


\begin{document}
\maketitle

\section{Introduction}\label{sec:intro}

Young supernova remnants (SNRs), microquasars, star formation regions (SFRs), and pulsar wind nebulae (PWNe) are prime candidates for Galactic
PeVatrons, accelerators of Galactic cosmic rays with maximum energies $E\sim 1$ PeV, which manifest as sources with a curved or broken 
hadronic $\gamma$-ray spectrum extending up to 100 TeV
\citep{2021Natur.594...33C, 2024ApJS..271...25C, devin2024pevfrontierstatusgammaray}.
The non-thermal high-energy (HE, $E > 100$ MeV) and very high-energy (VHE, $E > 100$ GeV) $\gamma$-ray emission from these sources should be a promising manifestation of
acceleration processes in Galactic cosmic-ray sources and is the main target for 
space-borne \citep{2023arXiv230712546B, 2026arXiv260222148B} and ground-based 
gamma-ray 
observatories \citep{2018A&A...612A...1H, 2026arXiv260200263A, 2024ApJS..271...25C}.
Until now, the Large High Altitude Air Shower Observatory (LHAASO) in China has detected dozens of Galactic PeVatrons as $\gamma$-ray sources in the energy range $E>0.1$ PeV (energy of $\gamma$-ray is about one-tenth of the cosmic ray energy) \citep{2021Natur.594...33C, 2024ApJS..271...25C}, but practically all of them are leptonic PeVatrons (like the Crab nebula) 
\citep{2021Natur.594...33C, 2024ApJS..271...25C}, and only discovered in 2024 source LHAASO J1912+1014u show the signatures of hadronic PeVatron \citep{2024ApJS..271...25C, 2026ApJ..1006...77M}.

Both hadronic and leptonic PeVatrons may be the result of a Supernova explosion with the formation of a millisecond magnetar \citep{2015JCAP...08..026K}. Magnetars are young neutron stars with extremely 
high surface magnetic fields $B_{\textrm{s}}\sim 10^{14}-10^{15}$ G 
\citep{1992ApJ...392L...9D, 1995MNRAS.275..255T, 2017ARA&A..55..261K, 2021ASSL..461...97E}.
The total magnetar luminosity is supported mainly by a decay of magnetar's magnetic energy
and considerably exceeds ordinary spin-down energy losses \citep{1995MNRAS.275..255T, 1996ApJ...473..322T}.

 The large magnetic fields of magnetars provide the necessary conditions for the formation of both leptonic and hadronic PeVatrons. For a newborn magnetar with the initial period of order of 1 millisecond
$P_{\textrm{i}}\sim 10^{-3} \textrm{ s}$ \citep{1995MNRAS.275..255T, 2017ApJ...841...14M}, its rotational kinetic energy
$E_{\textrm{rot}} \approx 2\times 10^{52}P_{\textrm{i,-3}}^{-2} \textrm{ erg}$ 
dissipates rapidly by magnetic dipole spin-down luminosity
$L_{\textrm{sd,i}} \approx 10^{49}B^2_{\textrm{s},15}P_{\textrm{i,-3}}^{-4} 
\textrm{ erg} \textrm{ s}^{-1}$
with a short initial spin-down time-scale
$t_{\textrm{sd,i}} = E_{\textrm{rot}}/L_{\textrm{sd,i}}
\approx 2\times 10^3 B^{-2}_{\textrm{s},15}P^{\,2}_{\textrm{i,-3}} \textrm{ s}$, for a neutron star with a moment of inertia $I=10^{45}$~g~cm$^2$ and a radius $R=10$~km \citep{1995MNRAS.275..255T, 2017ApJ...841...14M}.\footnote{Hereafter the physical quantities are expressed by $Q=10^{n}Q_{n}$ in cgs units, unless specified otherwise.}

Although a millisecond period is unlikely, a newborn magnetar can reach 
 a millisecond period due to the fallback accretion \citep{2020ApJ...903L..24L}. 

A rapid release of about $10^{52}$ erg of a fast-rotating magnetar in the form of a 
spherically symmetric collimated magnetar wind additionally energizes the 
Supernova ejecta up to $\sim 10^{52}$ erg of kinetic energy, turning it into a Hypernova
-- superluminous Supernova -- and later on into a $\sim 10^{52}$ erg Hypernova remnant (HNR)
\citep{1992ApJ...392L...9D, 2007MNRAS.380.1541B, 2019MNRAS.487.2215A}.
The jet-like collimated magnetar wind can drill into the ejecta shell and 
create a long gamma-ray burst (GRB) event
\citep{2003ApJ...589..871A,2009PhRvD..79j3001M,
2015PhR...561....1K,2018ApJ...869..155S,2019ApJ...871L..25P}.
 Owing to the presence of relativistic shock waves and reconnection processes in magnetar giant flares, magnetars are
promising accelerators of ultra high energy cosmic rays (UHECRs) \citep{2011ARA&A..49..119K,2019ApJ...878...34F}.
 Further, the magnetar SGR 1900+14 is a potential source of 
 a triplet, i.e. a group of three events, of extreme-energy cosmic rays (EECRs) with $E_{cr} > 10^{20}$~eV \citep{2026Galax..14...41Z}.

Other potential sources of Galactic PeVatrons are Galactic SFRs, which include young massive stellar clusters (YMSCs), OB associations,
and superbubbles \citep{2021MNRAS.504.6096M, 2021Univ....7..324C, 2023MNRAS.523.4015B, 2025A&A...695A.175M, 2024SciBu..69..449C, 2024arXiv240603555V}. In YMSCs, a joint action of powerful winds of massive stars
and SN ejecta creates a set of strong individual and large-scale termination shocks. As in the hypernova remnant considered above, cosmic rays are accelerated at these shocks by diffusive shock acceleration (DSA), which provides
a power-law type of CR spectra with spectral indexes $\Gamma_{cr} = 2.0-
2.3$ \citep{2021MNRAS.504.6096M, 2023MNRAS.523.4015B}. VHE $\gamma$-ray emission have been detected from some
YMSCs \citep{2021MNRAS.504.6096M} including magnetar SGR 1806-20-connected star cluster
Cl 1806-20 \citep{2018A&A...612A..11H, 2018A&A...612A...1H} and Westerlund 1 (Wd 1), the most massive young stellar cluster known in Galaxy \citep{2022A&A...664A.146N, 2025ApJ...992..213W, 2025NatCo..16.10820L}.

Both potential Galactic PeVatrons, connected with magnetar SGR1900+14 \citep{2009ApJ...707..844D, 2000ApJ...533L..17V} and YMSC W49A \citep{2005A&A...430..481H, 2013ApJ...779..121G, 2013ApJ...775...79Z, 2026arXiv260805453J}, overlap on the \lat{} unidentified extended
source \src{} (4FGL~J1908.6+0915e in the 4FGL-DR4 catalogue) \citep{2020ApJS..247...33A, 2022ApJS..260...53A, 2023arXiv230712546B}. In \citet{2022MNRAS.514..762H} (hereafter Paper 1), models of this source as a magnetar-connected Hypernova remnant (hadronic model) or magnetar wind nebula (leptonic model) have been elaborated. One of the weak points of the models was the large extent of the $\gamma$-ray source - about 250 pc for its radius of 0.6 deg at a distance of 12.5 kpc. In this work, we show that new data of the Fermi-LAT 16-year Source List \citep{2026arXiv260222148B} suggest splitting the source into two components - one in coincidence with magnetar SGR 1900+14 and the other with superbubble, adjacent to the YMSC W49A. In this new model, magnetar-connected Hypernova contributes as well to the LHAASO-detected source J1907+0913, while point-like Fermi-LAT source FL16Y J1910.2+0903 corresponds to the YMSC W49A itself and a new adjacent extended source to the W49A superbubble. 

The paper is structured as follows. In Section 2, we describe multiwavelength observations of the SGR 1900+14--W49A sky region from radio to $\gamma$-ray band. In Section 3, the results of modeling the multiband spectral energy distribution of the SGR 1900+14--W49A sky region are presented. The discussion and conclusions are presented in Section 4.

\section{Multiwavelength observations of the SGR 1900+14--W49A
sky region
\label{sec:region}}

\subsection {SGR 1900+14--W49A sky region in the gamma-ray band
\label{sec:gamma-band}}
\label{sec:mwl:tev}

Two \lat{} sources are located in the field studied here: the
extended source \src{}, modelled as a uniform disc of radius $0.6^\circ$ centred
on $(l,b)=(43.125^\circ,+0.430^\circ)$ and the point source FL16Y~J1910.2+0903, identified with the
star-forming region \wf{} \citep{2022MNRAS.514..762H}. All catalogue values
quoted below come from \texttt{gll\_psc\_v41.fit}, the 16~yr source
list built from P8R3 data spanning 2008~August~4 to 2024~August~2, in which
the sources carry the prefix \texttt{FL16Y}; this is the successor of
4FGL-DR4 \citep{2023arXiv230712546B}, itself the fourth data release of the 4FGL
catalogue \citep{2020ApJS..247...33A}. Using the most recent release matters
here because the disc-like FL16Y J1908.6+0915e is detected at only $2$--$3\sigma$ per energy band above 10~GeV,
which is precisely the range that any extrapolation towards the TeV data has to
rest on.

Neither source shows significant curvature ($\sigma_\mathrm{curv}=1.29$ and
$1.79$), so both are published, and used here, as power laws. The two indices
are close, $\Gamma=2.181\pm0.084$ for the disc against $2.304\pm0.074$ for the
point source.

\begin{figure}[H]
\includegraphics[width=\textwidth]{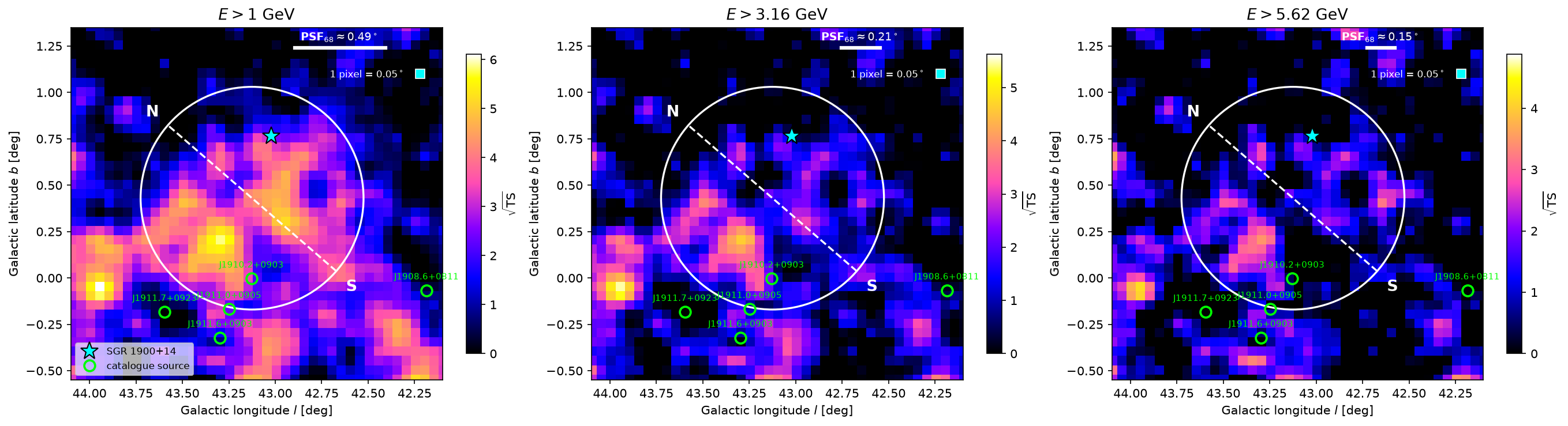}
\caption{$1$--$1000$~GeV significance maps for the region around
\magnetar{}--\wf{} from the 16~yr data set (Section~\ref{sec:morphology}, Table~\ref{tab:sed}); from left to right the maps use photons above $1$, $3.16$ and $5.62$~GeV, in Galactic coordinates. All catalogue sources, excluding \src{}, were considered and removed.
The green circles mark the positions of the catalogue sources, the cyan star
marks the position of \magnetar{}, the white circle represents the extension of
\src{} ($r=0.6^\circ$), and the white dashed line marks its division into the
northern (N, \magnetar{}) and southern (S, \wf{}) halves at
$\theta=+40.93^\circ$. The bar in each panel gives the $68$~per~cent containment
of the PSF and the cyan square one pixel ($0.05^\circ$). Counts, excess-counts and flux maps of the same region are shown in Figure~\ref{fig:maps} (Appendix~\ref{app:maps}).\label{fig:resid}}
\end{figure}

\begin{figure}[H]
 \centering
 \includegraphics[width=0.7\linewidth]
 {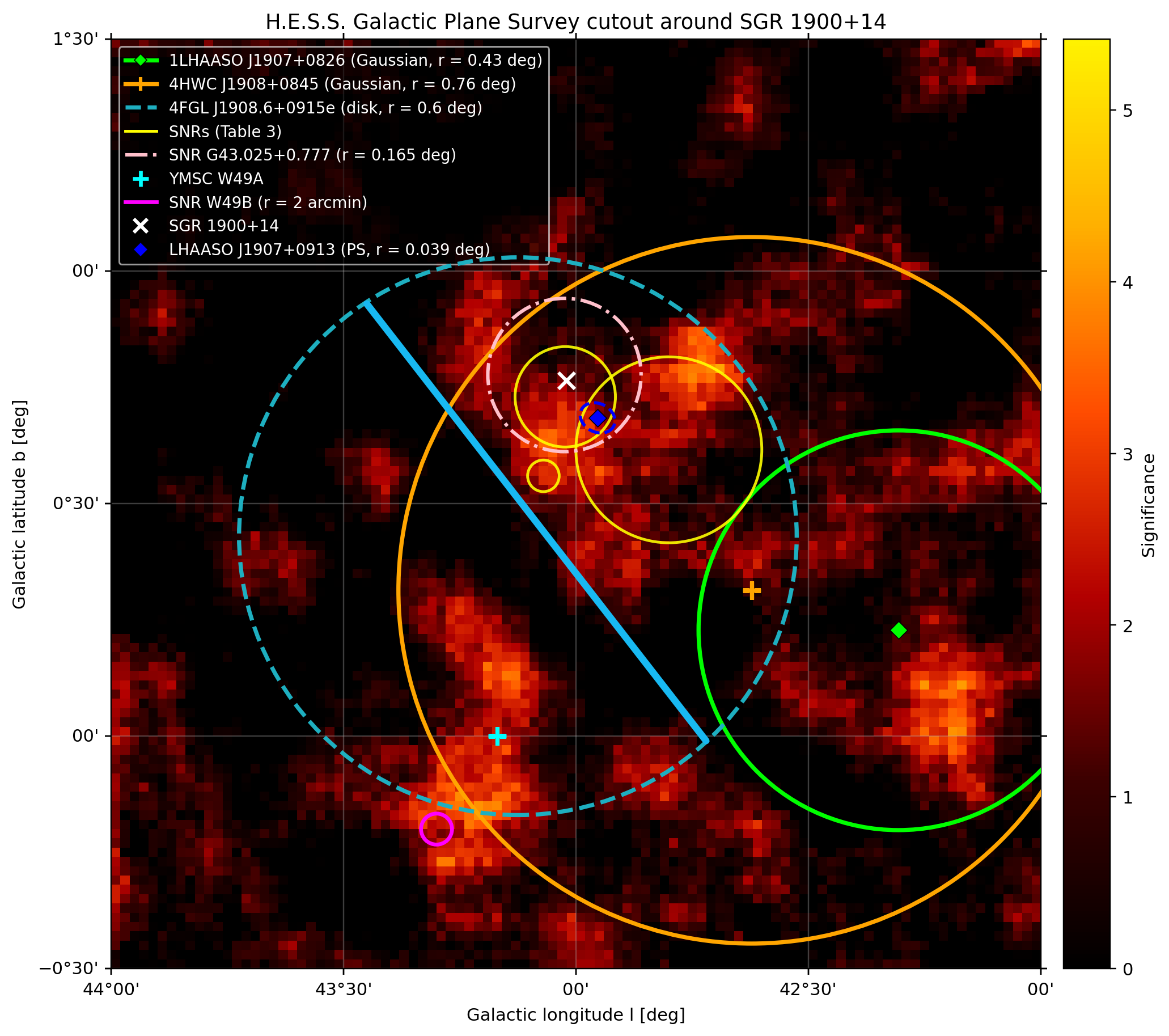}
 \caption{
 H.E.S.S. Galactic Plane Survey significance map of the SGR 1900+14--W49A sky region with the TeV sources and potential associated sources - SNRs and YMSC W49A. The background uses the public HGPS
 $0.1^{\circ}$ correlation-radius significance-map data \citep{2018A&A...612A...1H}. Yellow circles: the SNR G42.8+0.6 and the SNR candidates G43.023+0.73 and G43.07+0.56 of Table~\ref{tab:co-radio-sources}. 
 }
 \label{fig:sgr1900p14_hgps}
\end{figure}

\subsubsection{Morphology of \src{}}\label{sec:morphology}

The unidentified extended ($R=0.6$ deg) Fermi-LAT source \src{} contains the two cosmic ray accelerators - magnetar \magnetar{} in the north part of its disk and YMSC \wf{} in the south part (Figure~\ref{fig:resid}). To test whether the data allow for such source confusion, we performed a binned three-dimensional likelihood analysis with \texttt{fermipy} \citep{2017ICRC...35..824W}. We used P8R3 SOURCE events from 2008~August~4 to 2026~January~16 at 1--1000~GeV, with the four PSF event types as separate components of a joint likelihood, in a $15^\circ\times15^\circ$ region with $0.05^\circ$ pixels and eight energy bins per decade. The model contains the Galactic interstellar emission model \texttt{gll\_iem\_v07}, the isotropic templates of the four event types and the sources of the 16~yr source list; we repeated every test with 4FGL-DR4. The normalisations of all sources within $2^\circ$ of the disc and of the two diffuse components are free, and all models are fitted from the same optimised background model. The tested components have power-law spectra with free normalisation and index. We compare the models with the Akaike information criterion, ${\rm AIC}=2k-2\ln L$ \citep{1974ITAC...19..716A}, where $k$ is the number of free parameters of the tested components.

Two point sources with free positions, started at \magnetar{} and \wf{}, describe the data much worse than the catalogue disc ($\Delta{\rm AIC}=+236$ in the analysis of the full 17.5~yr data set), so the emission is extended.
We fitted two extended components with free positions, extensions and spectra. With the standard interstellar emission model these fits do not 
converge on the source: both components grow to the upper end of the extension scan, $R_{68}\approx1.5^\circ$, and are centered $0.8^\circ$ and $3^\circ$ from the catalogue position. Therefore, we replace the uniform 
disc by two half-discs, each with free normalisation and index, as a nested test with two additional parameters. It improves the fit by 
$1.8\sigma$ for the division of the disc along the line shown in Figure~\ref{fig:resid} ($\theta=+40.93^\circ$). In the 16~yr data set, analysed with the FL16Y configuration and reproducing the published flux of the disc, the improvement is $1.2\sigma$; this data set is used for the spectra of the two halves (Table~\ref{tab:sed}).

The catalogue spectrum of the point source FL16Y~J1910.2+0903 inside the southern half-disc, which is modelled separately in Section~\ref{sec:south:models}, is given in Table~\ref{tab:ps-sed}.

\begin{table}[H]
\caption{\lat{} spectra of the northern (\magnetar{}) and southern (\wf{}) half-discs of \src{} for the division of Figure~\ref{fig:resid} ($\theta=+40.93^\circ$), in the five catalogue bands above 1~GeV, compared with the catalogue spectrum of the whole disc \citep{2026arXiv260222148B}. $E^2\,dN/dE$ in $10^{-9}$~GeV~cm$^{-2}$~s$^{-1}$ with $1\sigma$ errors; $95$~per cent upper limits where ${\rm TS}<4$; $\sqrt{\rm TS}$ is the significance in the band.\label{tab:sed}}
\small\begin{tabularx}{\textwidth}{lCCCCCCC}
\toprule
 & & \multicolumn{2}{c}{\textbf{Northern half-disc}} & \multicolumn{2}{c}{\textbf{Southern half-disc}} & \multicolumn{2}{c}{\textbf{Whole disc (catalogue)}} \\
\cmidrule(lr){3-4}\cmidrule(lr){5-6}\cmidrule(lr){7-8}
\textbf{Band} & \boldmath{}$E_\mathrm{ref}$ \textbf{[GeV]} & \boldmath{}$E^2\,dN/dE$ & \boldmath{}$\sqrt{\rm TS}$ & \boldmath{}$E^2\,dN/dE$ & \boldmath{}$\sqrt{\rm TS}$ & \boldmath{}$E^2\,dN/dE$ & \boldmath{}$\sqrt{\rm TS}$ \\
\midrule
1--3 GeV & 1.732 & $1.27\pm0.41$ & 3.2 & $1.21\pm0.50$ & 2.5 & $3.08\pm0.61$ & 5.1 \\
3--10 GeV & 5.477 & $<0.59$ & 0.0 & $1.90\pm0.43$ & 4.6 & $1.71\pm0.49$ & 3.6 \\
10--30 GeV & 17.321 & $<1.13$ & 1.3 & $<0.92$ & 0.4 & $1.20\pm0.56$ & 2.3 \\
30--100 GeV & 54.772 & $<0.85$ & 0.0 & $<1.15$ & 0.4 & $<1.80$ & 0.9 \\
100--1000 GeV & 316.228 & $1.16\pm0.62$ & 2.3 & $1.31\pm0.76$ & 2.0 & $2.26\pm0.94$ & 3.0 \\
\bottomrule
\end{tabularx}
\noindent{\footnotesize{Notes: $E_\mathrm{ref}=\sqrt{E_\mathrm{min}E_\mathrm{max}}$. The half-disc spectra are from our binned likelihood analysis of the 16~yr data set (2008~August--2024~August) above 1~GeV with the analysis configuration of the 16~yr source list (energy- and PSF-type components, rescaled interstellar emission model), in which the whole disc reproduces its catalogue flux and TS within the errors. In each band the normalisations of both half-discs and of the Galactic interstellar emission are free and the other sources are fixed at the global fit; the spectral index within a band is fixed at the catalogue value of the disc, $2.18$. Errors are statistical. The two half-discs add up to the fitted flux of the whole disc within $3$~per cent. The catalogue columns give the published band fluxes of the disc and their significance; the upper limit is $F+2\,\sigma_\mathrm{up}$.}}
\end{table}

\begin{table}[H]
\caption{\textit{Fermi}-LAT catalogue spectrum of the point source FL16Y~J1910.2+0903, the cloud core of \wf{}, in the eight bands of the FL16Y source list \citep{2026arXiv260222148B}. $E^2\,dN/dE$ in $10^{-9}$~GeV~cm$^{-2}$~s$^{-1}$ with $1\sigma$ errors; $95$~per cent upper limits where ${\rm TS}<1$ (the convention of the catalogue); $\sqrt{\rm TS}$ is the significance of the source in the band.\label{tab:ps-sed}}
\small\begin{tabularx}{\textwidth}{lCCC}
\toprule {\bfseries\boldmath Band} & {\bfseries\boldmath $E_\mathrm{ref}$ [GeV]} & {\bfseries\boldmath $E^2\,dN/dE$} & {\bfseries\boldmath $\sqrt{\rm TS}$} \\
\midrule
$50$--$100$ MeV & $0.0707$ & $<20.6$ & 0.3 \\
$100$--$300$ MeV & $0.173$ & $<14.3$ & 0.4 \\
$0.3$--$1$ GeV & $0.548$ & $4.80\pm2.66$ & 1.8 \\
$1$--$3$ GeV & $1.732$ & $2.31\pm0.71$ & 3.4 \\
$3$--$10$ GeV & $5.477$ & $2.72\pm0.35$ & 9.3 \\
$10$--$30$ GeV & $17.32$ & $0.52\pm0.22$ & 3.2 \\
$30$--$100$ GeV & $54.77$ & $0.61\pm0.29$ & 3.4 \\
$100$--$1000$ GeV & $316.2$ & $1.50\pm0.61$ & 4.4 \\
\bottomrule
\end{tabularx}
\noindent{\footnotesize{Notes: $E_\mathrm{ref}=\sqrt{E_\mathrm{min}E_\mathrm{max}}$. Values from the FL16Y source list; the errors are the means of its asymmetric $1\sigma$ errors, as used in the fit of Table~\ref{tab:south-models}. The catalogue describes the source by a power law with photon index $2.30\pm0.07$ and $N_0=(3.32\pm0.39)\times10^{-14}$~MeV$^{-1}$~cm$^{-2}$~s$^{-1}$ at $7.23$~GeV (blue band in Figure~\ref{fig:south_pointsource_hadronic}); its significance is $10.2\sigma$.}}
\end{table}

\label{sec:lhaaso0913}

Three VHE sources lie in or close to the disc. For each of them we quote below
only the quantities that enter the modelling of Section~\ref{sec:modelling};
the complete catalogue entries are given in the references cited.

\textit{4HWC~J1908+0845.} The source is extended, with a symmetric
two-dimensional Gaussian morphology of $\sigma_{\mathrm{ext}}=0.76^{+0.08}_{-0.07}$~deg
centred at $l=42.62^\circ$, $b=+0.31^\circ$, so it covers both halves of the
disc and does not separate them. Its spectrum is a simple power law with
$N_0=\left(11.5^{+1.39}_{-1.28}\right)\times10^{-13}$~TeV$^{-1}$\,cm$^{-2}$\,s$^{-1}$
at $E_0=2$~TeV and $\Gamma=2.72\pm0.03$ \citep{2025icrc.confE.665G}. We use this
spectrum, integrated over the full Gaussian region, as an upper limit on the
southern components (Section~\ref{sec:south:models}), and we draw it as a
butterfly in the SED figures, i.e. as the $1\sigma$ confidence band of the power-law spectrum (also called a bow-tie).

\textit{1LHAASO~J1907+0826.} The source is detected by WCDA in the $1$--$25$~TeV
range with a Gaussian extension $r_{39}=(0.43\pm0.08)^\circ$, and lies
$0.92^\circ$ from the centre of the disc, that is, outside it. Its WCDA spectrum
is a power law,
$N_0=(1.34\pm0.29)\times10^{-13}$~TeV$^{-1}$\,cm$^{-2}$\,s$^{-1}$ at
$E_0=3$~TeV with $\Gamma=2.62\pm0.14$ \citep{2024ApJS..271...25C}. It is not
significantly detected by KM2A above $25$~TeV, and the KM2A upper limit
indicates a possible steepening or cut-off, so we do not extrapolate the power
law beyond $25$~TeV. Being a separate object outside the disc, it enters the
modelling only as the butterfly drawn for its full Gaussian region in the SED
figures. The quoted uncertainties are statistical, and the catalogue gives no
covariance between $N_0$ and $\Gamma$.

\textit{\lhaaso{}.} The point source \lhaaso{} ($10.6\sigma$~\citep{2026arXiv260625054C}) lies $6'$ from the magnetar, inside the $9.9'$ radio shell of G43.025+0.777, and is the first significant TeV detection at the position of the northern component. No spectrum of it has been published. In an earlier LHAASO analysis of the field \citep{2024arXiv241008988C} it is detected by WCDA at $8.0\sigma$ in $1$--$25$~TeV, and its KM2A counterpart LHAASO~J1907+0910 at $5.8\sigma$ above $25$~TeV. We estimate its flux from these significances, with the relation between flux and significance calibrated on the sources of the 1LHAASO catalogue \citep{2024ApJS..271...25C} at similar declinations and scaled to the catalogue exposure. This gives $E^2\,{\rm d}N/{\rm d}E\approx6\times10^{-10}$~$\gevcms$ at $3$~TeV and $\approx5.2\times10^{-11}$~$\gevcms$ at $50$~TeV, with errors of a factor of $1.5$ and $2$ that include the scatter of the calibration, the unknown photon index and the statistical error of the significance. The two estimates imply a photon index of $\approx2.9$ between $3$ and $50$~TeV and are the only two TeV points used as detections in the fit, in the model of the northern component (Section~\ref{sec:xray:model}, Figure~\ref{fig:north_sed}). The WCDA and KM2A positions agree within $0.1^\circ$, much less than the angular resolution of the detectors, and the LHAASO collaboration describes the emission from 1 to above 100~TeV with a single point source \citep{2026arXiv260625054C}, so the two points belong to one source. The KM2A significance depends on the fit of the region: $5.8\sigma$ above 25~TeV in the fit of the GRS~1915+105 field and $4.1\sigma$ at 25--100~TeV in the fit of the MGRO~J1908+06 field \citep{2024arXiv241008988C}. Since a significance does not correspond to a unique flux, the two estimates are approximate. They amount to 17 and 11 per cent of the total flux of 4HWC~J1908+0845 at 3 and 50~TeV. The dependence of our results on them is discussed in Section~\ref{sec:xray:model}.

 \subsection {SGR 1900+14--W49A sky region in the radio-optical band
\label{sec:radio-band}}
\begin{figure}[H]
 \centering
 \includegraphics[width=\textwidth]{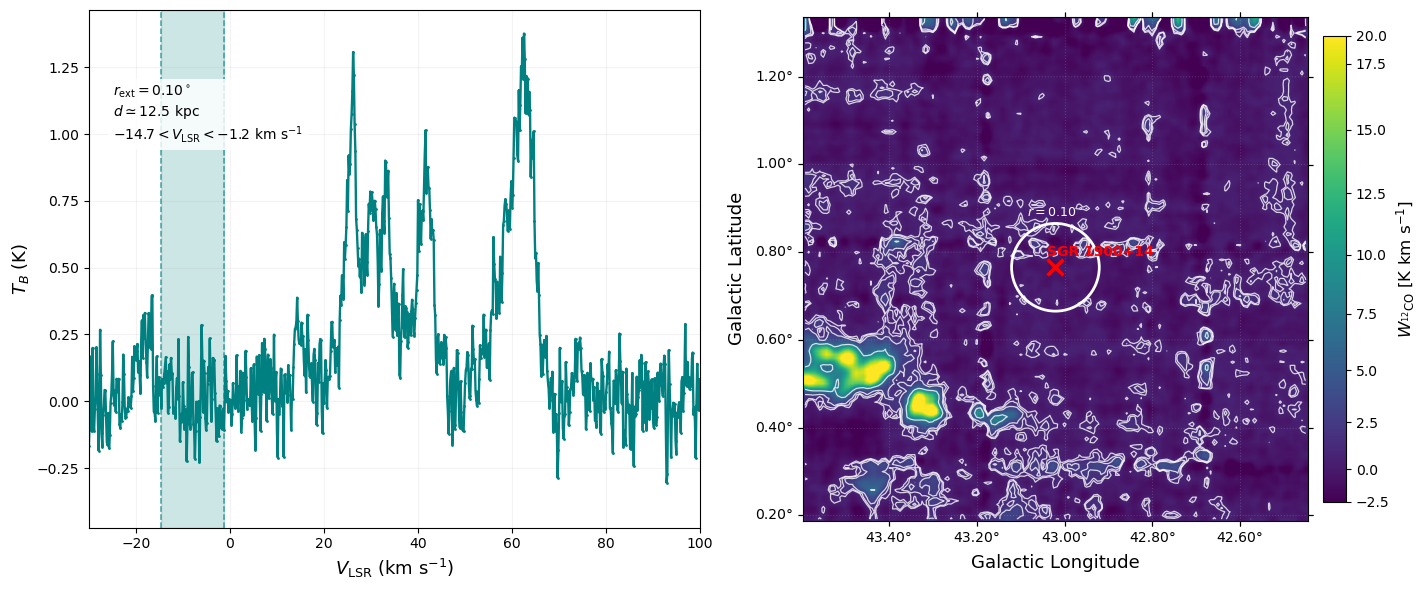}
 \caption{
 $^{12}\mathrm{CO}$ molecular-gas environment of SGR~1900+14.
 The left panel shows the spectrum extracted within a radius of
 $0.10^{\circ}$. The shaded interval,
 $-14.7 < V_{\mathrm{LSR}} < -1.2~\mathrm{km\,s^{-1}}$,
 corresponds to the velocity range used to construct the
 integrated-intensity map and to an adopted kinematic distance of
 approximately $12.5~\mathrm{kpc}$. The right panel shows the
 corresponding $^{12}\mathrm{CO}$ integrated-intensity map.
 The white circle marks the spectral-extraction region, and the
 red cross indicates the position of SGR~1900+14.
 The $^{12}\mathrm{CO}(J=1\text{--}0)$ data are taken from the
 Milky Way Imaging Scroll Painting (MWISP) survey
 \citep{2019ApJS..240....9S,Yang2026MWISP}.
 }
 \label{fig:co-sgr1900}
\end{figure}

In SGR 1900+14--W49A sky region there are one SNR G42.8+0.6 and three recently discovered SNR candidates G43.023+0.73, G43.07+0.56 and G43.025+0.777 
(Figure~\ref{fig:sgr1900p14_hgps} and Table~\ref{tab:co-radio-sources}). SNRC G43.025+0.777, whose center coincides with the magnetar position, may be a magnetar-connected Supernova/Hypernova remnant or 
magnetar wind nebula. Till now, there are no SNR or PWN in W49A. Figure~\ref{fig:co-sgr1900} does not show the presence of considerable molecular clouds in the magnetar vicinity. 
Meantime, in W49 vicinity, there are no SNR or PWN, but the well-known massive molecular cloud structures are presented \citep{2026arXiv260805453J}.

\begin{table}[H]
\caption{CO-line velocities, kinematic distances and radio flux densities of the sources in the northern half of \src{}. $v_{\rm peak}$ is the velocity of the strongest $^{12}$CO(1--0) component in the aperture of each source, from the MWISP survey \citep{2019ApJS..240....9S}. $d_{\rm near}$ and $d_{\rm far}$ are the two kinematic distances of this gas for a flat rotation curve with $R_0=8.15$~kpc and $\Theta_0=236$~km~s$^{-1}$ \citep{2019ApJ...885..131R}. They refer to the emitting gas, not to established distances of the radio sources. Flux densities are integrated over each source.\label{tab:co-radio-sources}}
\begin{tabularx}{\textwidth}{lCCCCCc}
\toprule
\textbf{Source} & \boldmath{$v_{\rm peak}$} & \boldmath{$d_{\rm near}$} & \boldmath{$d_{\rm far}$} & \boldmath{$\nu$} & \boldmath{$F_\nu$} & \textbf{Ref.} \\
 & ($\mathrm{km\,s^{-1}}$) & (kpc) & (kpc) & (GHz) & (mJy) & \\
\midrule
G42.8+0.6 & 41.6 & 2.63 & 9.33 & 1.42 & $2400\pm600$ & {\citep{1987A&AS...69..403F}} \\
G43.023+0.73 & 62.7 & 4.09 & 7.83 & 5.8 & $7.2\pm3.9$ & {\citep{2021A&A...651A..86D}} \\
G43.07+0.56 & 63.2 & 4.14 & 7.77 & 5.8 & $32.2\pm1.8$ & {\citep{2021A&A...651A..86D}} \\
G43.025+0.777 & 60.8 & 3.94 & 7.98 & 1.28 & $477\pm13$$^{a}$ & {\citep{2025A&A...695A.144B}} \\
\bottomrule
\end{tabularx}
\noindent{\footnotesize{Note: $^{a}$Flux density of the shell after removal of the $50$ compact sources inside it (column \texttt{FluxnoPS} of the SMGPS extended-source catalogue); with these sources included the catalogue gives $531\pm13$~mJy.}}
\end{table}

\begin{figure}[H]
 \centering
 \includegraphics[width=\textwidth]{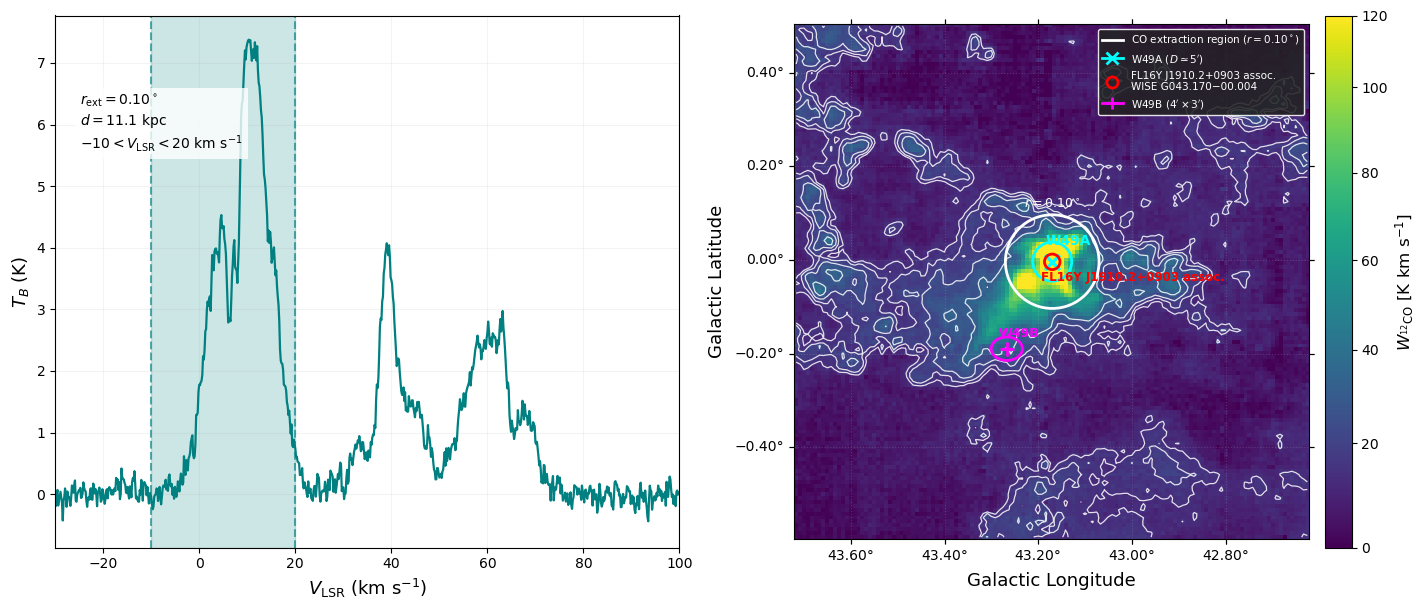}
 \caption{
 $^{12}\mathrm{CO}$ molecular-gas environment of W49A and
 FL16Y~J1910.2+0903. The left panel shows the spectrum extracted
 within a radius of $0.10^{\circ}$. The shaded interval,
 $-10 < V_{\mathrm{LSR}} < 20~\mathrm{km\,s^{-1}}$,
 is used to construct the integrated-intensity map and corresponds
 to an adopted kinematic distance of approximately
 $11.1~\mathrm{kpc}$. The right panel shows the integrated
 $^{12}\mathrm{CO}$ emission, together with the positions or regions
 associated with W49A, FL16Y~J1910.2+0903,
 the WISE H\,{\sc ii} region, and the supernova remnant W49B.
 The $^{12}\mathrm{CO}(J=1\text{--}0)$ data are taken from the
 Milky Way Imaging Scroll Painting (MWISP) survey
 \citep{2019ApJS..240....9S,Yang2026MWISP}.
 }
 \label{fig:co-w49a}
\end{figure}

The radio emission of \wf{} is thermal. The SMGPS catalogue of extended sources \citep{2025A&A...695A.144B} lists the H\,{\sc ii} complex G043.180$-$00.016 ($r=5.9'$) at $43.7$~Jy at $1.28$~GHz. This is the WISE region associated with the \lat{} source. In the southern half-disc the catalogue has $46$ sources with $75$~Jy in total, most of them H\,{\sc ii} regions of the complex. We use these fluxes only as upper limits on synchrotron emission, $43.7$~Jy for the point source and $75$~Jy for the half-disc.

 It is important to discuss the Hypernova signatures of magnetar-connected Supernova in the optical band. As analyzed in detail in Paper 1, a sky 
 position, distance, and age of the magnetar SGR 1900+14 agree with the
 4 BCE "po star"-a transient ($\sim$ 1 month) immovable (i.e. not moving with respect to the stars, unlike a comet) source visible by naked eye ($\sim$ 5 mag)- from ancient Chinese record dated 4 BC April 24 
 \citep{2002ApJ...569L..43W, 2005JHA....36..217S}. 4 BC April 24 also corresponds to the Star of Bethlehem. For the visual extinction $A_V = 12.9\pm 0.5$ mag \citep{2009ApJ...707..844D} SN explosion would have been optically observable only when the absolute magnitude of a supernova near the maximum light was of order 
 $M\approx -23$ mag. An improved SGR 1900+14 age estimate $\tau_{\textrm{c}}\approx 2.4$ kyr from new $P$ and 
 $\dot P$ data \citep{2019PASJ...71...90T} agrees with the
 "po star" age, taking into account the uncertainty of 
 the spin-down rate $\dot P$ determination \citep{2019PASJ...71...90T}.

\subsection {SGR 1900+14--W49A sky region in the X-ray band
\label{sec:X-ray-band}}
\label{sec:xray}

\subsubsection{Archival X-ray Observations of \magnetar{}}
\magnetar{} has been observed by all major X-ray observatories since its identification as a soft $\gamma$-ray repeater \citep{1979SvAL....5..343M, 1999ApJ...510L.115K}. A \textit{Chandra} observation made soon after the 2001 April reactivation revealed a dust-scattering halo extending to $\gtrsim100''$ from the source \citep{2001ApJ...558L..47K}. Three further \textit{Chandra}/ACIS pointings over five years set a $3\sigma$ upper limit of $70$~mas~yr$^{-1}$ on the proper motion of the magnetar \citep{2009ApJ...692..158D}. At this speed the magnetar could not have moved from the centre of the nearby SNR G42.8+0.6 \citep{2019JApA...40...36G} to its present position within the age of the remnant, which argues against an association between the two.

\textit{XMM-Newton} observed the magnetar in 2005 September and again in 2006 April, as a target of opportunity after new burst activity. \citet{2006ApJ...653.1423M} found a stable spectrum, a blackbody plus a power law ($kT\approx0.47$~keV, $\Gamma\approx1.9$, $N_{\rm H}\approx2.1\times10^{22}$~cm$^{-2}$), with no emission or absorption lines down to equivalent widths of $50$--$200$~eV. \citet{2019PASJ...71...90T} extended the spectrum to $70$~keV with simultaneous \textit{XMM-Newton} and \textit{NuSTAR} observations in 2016 October ($kT=0.52$~keV, $\Gamma=1.21$, $N_{\rm H}=1.96\times10^{22}$~cm$^{-2}$). All these studies treated the magnetar as a point source, and no X-ray counterpart of an SNR or PWN connected with the magnetar has been reported.

\subsubsection{Stacked \textit{XMM-Newton} and \textit{Chandra} Images: No Counterpart of G43.025+0.777}
The seven \textit{XMM-Newton} pointings of 2005--2016 give a combined EPIC exposure of $\approx160$~ks. The MeerKAT shell G43.025+0.777 is a candidate SNR with the radio centre at RA~$=286.801^\circ$, Dec~$=+9.33^\circ$ ($l=43.025^\circ$, $b=0.777^\circ$) and a radius $r=0.165^\circ=9.9'$. All pointings are centred on the magnetar to within a few arcminutes, so the shell lies almost entirely within the EPIC pn field of view in every observation. The archival data can therefore be stacked to search for an X-ray counterpart of the shell.

\begin{figure}[H]
 \centering
 \includegraphics[width=0.75\textwidth]{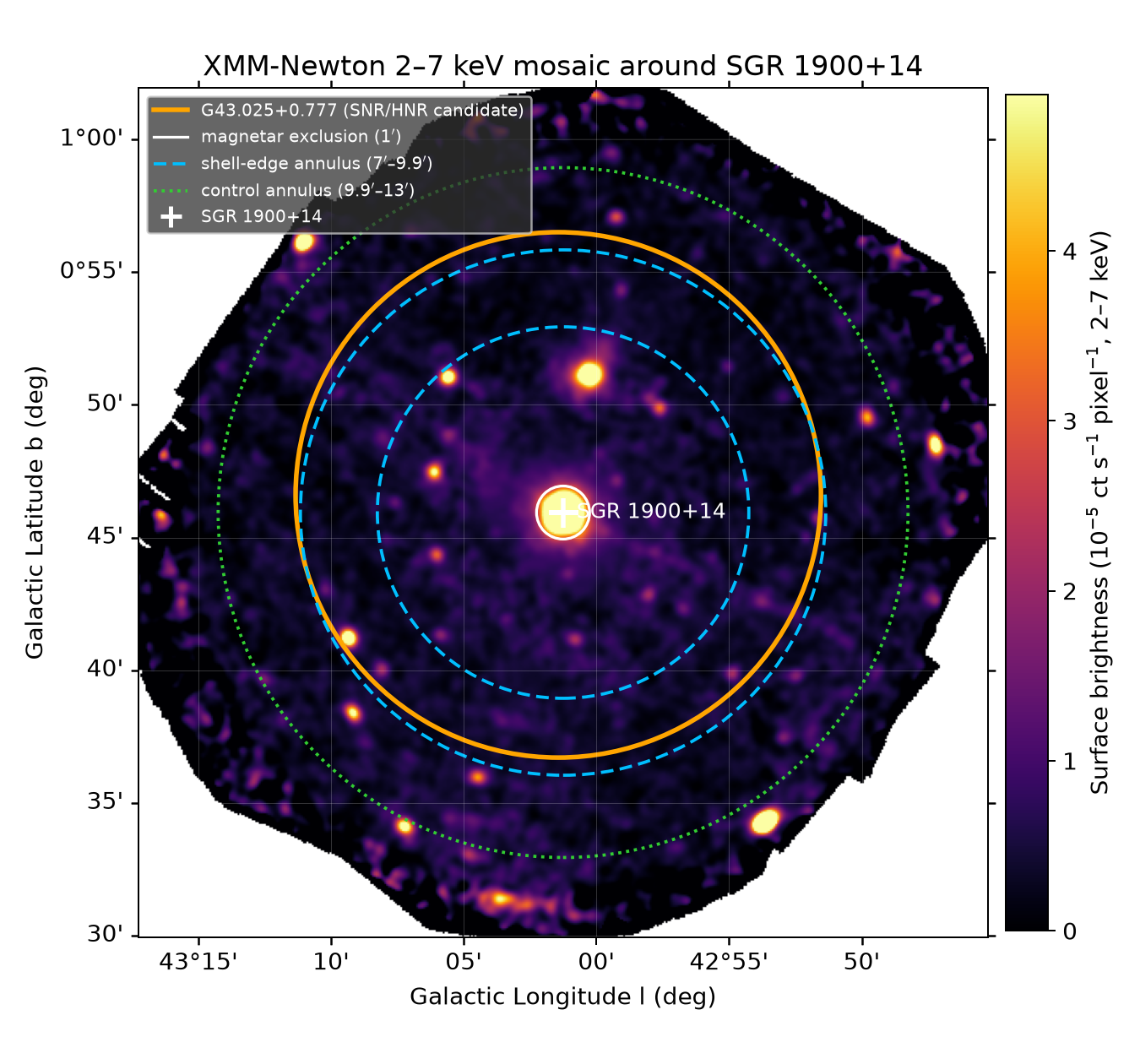}
 \caption{Stacked \textit{XMM-Newton} EPIC map of the \magnetar{} field in the $2$--$7$~keV band, corrected for background and exposure, in Galactic coordinates; sixteen datasets from six ObsIDs (Section~\ref{sec:xray}). The irregular boundary is the overlap of the fields of view of the six pointings at different roll angles. Orange circle: the radio shell G43.025+0.777 ($r=9.9'$, centred on its radio centre). White circle: the $1'$ region excluded around the magnetar (white cross). Dashed cyan and dotted green circles, both centred on the magnetar ($42''$ from the radio centre): the annulus at the shell edge ($7'$--$9.9'$) and the control annulus next to it ($9.9'$--$13'$) used in the significance test. The bright point sources in the field are unrelated to G43.025+0.777.}
 \label{fig:xray_galactic_map}
\end{figure}

We reprocessed all seven ObsIDs with SAS~22.1 (\texttt{epchain}/\texttt{emchain}) and removed soft-proton flares with \texttt{espfilt}. The short observation 0410580201 was excluded because residual flares keep its $2$--$7$~keV count rate $5$--$6$ times higher than in the other pointings after screening. In the shortest pointing, 0506430201, pn was not operating and MOS2 did not pass the flare screening. For the remaining sixteen EPIC datasets (six ObsIDs with up to three detectors each) we followed the ESAS procedure \citep{2008A&A...478..615S, 2008A&A...478..575K} and modelled the quiescent particle background with \texttt{mosback} and \texttt{pnback}. This background is nearly uniform over the detector, so we spread its predicted total evenly over the field of view of each dataset. The sixteen sets of counts, exposure and background were then combined on a common grid centred on the magnetar (Figure~\ref{fig:xray_galactic_map}).

The background-subtracted surface brightness inside G43.025+0.777, with the magnetar excluded (Figure~\ref{fig:xray_galactic_map}), is $34\pm3$~per cent higher than in an annulus at $12'$--$18'$ from the radio centre of the shell (Poisson errors only). The radial profile around the magnetar, however, falls smoothly from the edge of the excluded region to the field background at $\sim10'$ and shows no increase at the shell radius of $9.9'$. This is the dust-scattering halo of \citet{2001ApJ...558L..47K}, which our deeper stack follows to fainter levels than the $\gtrsim100''$ seen in the discovery observation. It does not follow the radio contour. We then compared a narrow annulus at the shell boundary ($7'$--$9.9'$ from the magnetar, where the halo is already faint) with a control annulus just outside it ($9.9'$--$13'$), with all detected point sources masked to $25''$. The difference is $6\pm2$~per cent, or $885\pm337$ counts, an excess of only $2.6\sigma$.

We also reprocessed the two \textit{Chandra}/ACIS-I pointings in which the halo was discovered (ObsIDs 1954 and 6731, 2001 and 2006, $\approx55$~ks in total). At sub-arcsecond resolution they show the same smooth decline out to $\sim8'$. At the position of the young radio pulsar PSR~J1907+0918 \citep{2000ApJ...545..385L}, $2'$ from the magnetar, there is no excess, so there is no bright X-ray PWN around it.

\subsubsection{Thermal and Nonthermal Limits on the Shell}\label{sec:xray:limits}
We compare the X-ray limit with the hypernova model of Paper~1. In that model a remnant with $E_{\rm HNR}\sim10^{52}$~erg has expanded for $t\approx2$~kyr inside a bubble of low density ($n_w\sim10^{-3}$~cm$^{-3}$) blown by the winds of the massive stars of the progenitor cluster. Its shock has reached a radius $R_{\rm HNR}\approx35$~pc with a velocity $V_{\rm HNR}\approx7\times10^3$~km~s$^{-1}$ and is about to reach the dense shell ($n_{\rm sh}\sim10$~cm$^{-3}$) formed by the winds at $R_{\rm sh}\approx40$~pc. The predicted shock radius agrees with the radius of G43.025+0.777, which was not known at the time, and with that of the VHE source HOTS~J1907+091 ($37\pm9$~pc in the same model). Behind a strong shock with velocity $V_{\rm HNR}$ the wind gas is heated to $kT\approx(3/16)\mu m_p V_{\rm HNR}^2\approx59$~keV. For such a plasma, the count limit at the shell edge converted with the EPIC response gives a $3\sigma$ upper limit on the absorbed $2$--$7$~keV flux of $5.1\times10^{-14}$~$\ergcms$ (a luminosity of $9.6\times10^{32}$~erg~s$^{-1}$ at $d=12.5$~kpc). For such hot gas the exponential factor $\exp(-E/kT)$ of the bremsstrahlung spectrum changes by only $8$~per~cent between $2$ and $7$~keV, so the shape of the spectrum in the band, and hence the limit hardly depends on the exact temperature. The emitting gas is the shocked wind, since the shock has not yet reached the outer shell. We compare the limit with the emission measure of a Sedov blast wave of radius $R_{\rm HNR}$ in a medium of density $n_w$, of which $64$~per cent projects into the $7'$--$9.9'$ annulus. For $n_w=10^{-3}$~cm$^{-3}$ the predicted flux is $0.07$ of the limit, too faint to be tested with these data, and the limit on the density of the bubble is $n_w\lesssim4\times10^{-3}$~cm$^{-3}$.

For a synchrotron rim the relevant limit is the one for a power law, $4.6\times10^{-14}$~$\ergcms$ ($\Gamma=2$, $N_{\rm H}=1.9\times10^{22}$~cm$^{-2}$), or $5.3\times10^{-14}$~$\ergcms$ unabsorbed. It is about $0.015$ of the only bound available before, used in Paper~1, the total unabsorbed $2$--$10$~keV flux of the magnetar itself, $F_{2-10\,{\rm keV}}=3.5\times10^{-12}$~$\ergcms$ (power law $1.78\times10^{-12}$~$\ergcms$ plus blackbody $1.7\times10^{-12}$~$\ergcms$ in this band)~\citep{2019PASJ...71...90T}. In the models of Section~\ref{sec:xray:model} this limit constrains the magnetic field and the electron spectrum directly.

\subsubsection{X-Rays of the \wf{} Complex}
The southern half-disc as a whole has not been observed in X-rays, and the archival observations cover only the W49 complex. There are two \textit{Chandra}/ACIS-I observations centred on \wf{} (ObsIDs 5893 and 6355, $96$~ks after cleaning), in which \citet{2006ApJ...653..409T} detected hard and strongly absorbed emission from the ultracompact H\,{\sc ii} regions ($kT\approx7$~keV, $N_{\rm H}\approx5\times10^{23}$~cm$^{-2}$), and five \textit{XMM-Newton} observations of W49B ($\approx0.47$~Ms summed over the three cameras), with \wf{} at the western edge of the field. We reprocessed all of them as the magnetar data above. Within $2'$ of \wf{}, which is the region of the point source, the total background-subtracted $2$--$7$~keV emission is $1609\pm83$ counts. For $N_{\rm H}=5\times10^{22}$~cm$^{-2}$ this is $\approx5.5\times10^{-13}$~$\ergcms$ unabsorbed, mostly thermal emission. We use $6.3\times10^{-13}$~$\ergcms$ (the total plus $3\sigma$) as an upper limit on synchrotron emission from the point source. Outside the complex the background-subtracted surface brightness is flat from $3'$ to $8'$ from \wf{}. The annulus at $3'$--$5'$ ($10$--$16$~pc) shows no excess over the annulus at $5'$--$8'$ ($4\pm159$~counts), which gives a $3\sigma$ upper limit on diffuse emission of $1.9\times10^{-13}$~$\ergcms$ unabsorbed ($L\lesssim3\times10^{33}$~erg~s$^{-1}$), or $3.7\times10^{-15}$~$\ergcms$~arcmin$^{-2}$. The adopted absorbing column density, $N_{\rm H}=5\times10^{22}$~cm$^{-2}$, is at the lower end of the values measured in X-rays toward W49B at the same distance, $5.2\times10^{22}$~cm$^{-2}$ for the whole remnant \citep{2007ApJ...654..938K} and $4$--$12\times10^{22}$~cm$^{-2}$ across it \citep{2013ApJ...764...50L}. For $10^{23}$~cm$^{-2}$ the limit is $1.4$ times higher. The \textit{XMM-Newton} data also show no excess within $8'$, but cannot be used at larger radii because W49B dominates the only side of \wf{} which they cover. For the half-disc we scale the annulus limit to its area and obtain $7.5\times10^{-12}$~$\ergcms$. This assumes a uniform surface brightness, as \citet{2026ApJ..1006...77M} did for LHAASO~J1912+1014u, although the annulus covers only $2.5$~per cent of the half-disc.

Neither limit constrains the magnetic fields assumed in the models of Section~\ref{sec:south:models}, $70\,\mu$G in the cloud core and $10\,\mu$G in the superbubble. For their electrons, cut off at $10$~TeV, the X-ray limits give $B\lesssim200\,\mu$G for the point source and $B\lesssim70\,\mu$G in the superbubble, about 3 and 7 times the assumed values.

\section{Modeling the multiband spectral energy distribution of the SGR 1900+14--W49A sky region
\label{sec:modelling}}

\subsection {Modeling the SGR 1900+14 sky region
\label{sec:modeling-magnetar}}
\label{sec:xray:model}

We model the SED of the northern half-disc with \textsc{naima} \citep{2015ICRC...34..922Z} in the two scenarios of Paper 1: a hypernova remnant, in which the $\gamma$-rays are mostly hadronic, and a magnetar wind nebula, in which they are inverse Compton emission of electrons.

 The $\gamma$-ray data are the two detected bands and the three upper limits of the northern half-disc (Table~\ref{tab:sed}).

We also use the two LHAASO flux estimates at $3$ and $50$~TeV for the point source inside the radio shell (Section~\ref{sec:lhaaso0913}).

\begin{figure}[H]
 \centering
 \includegraphics[width=\textwidth]{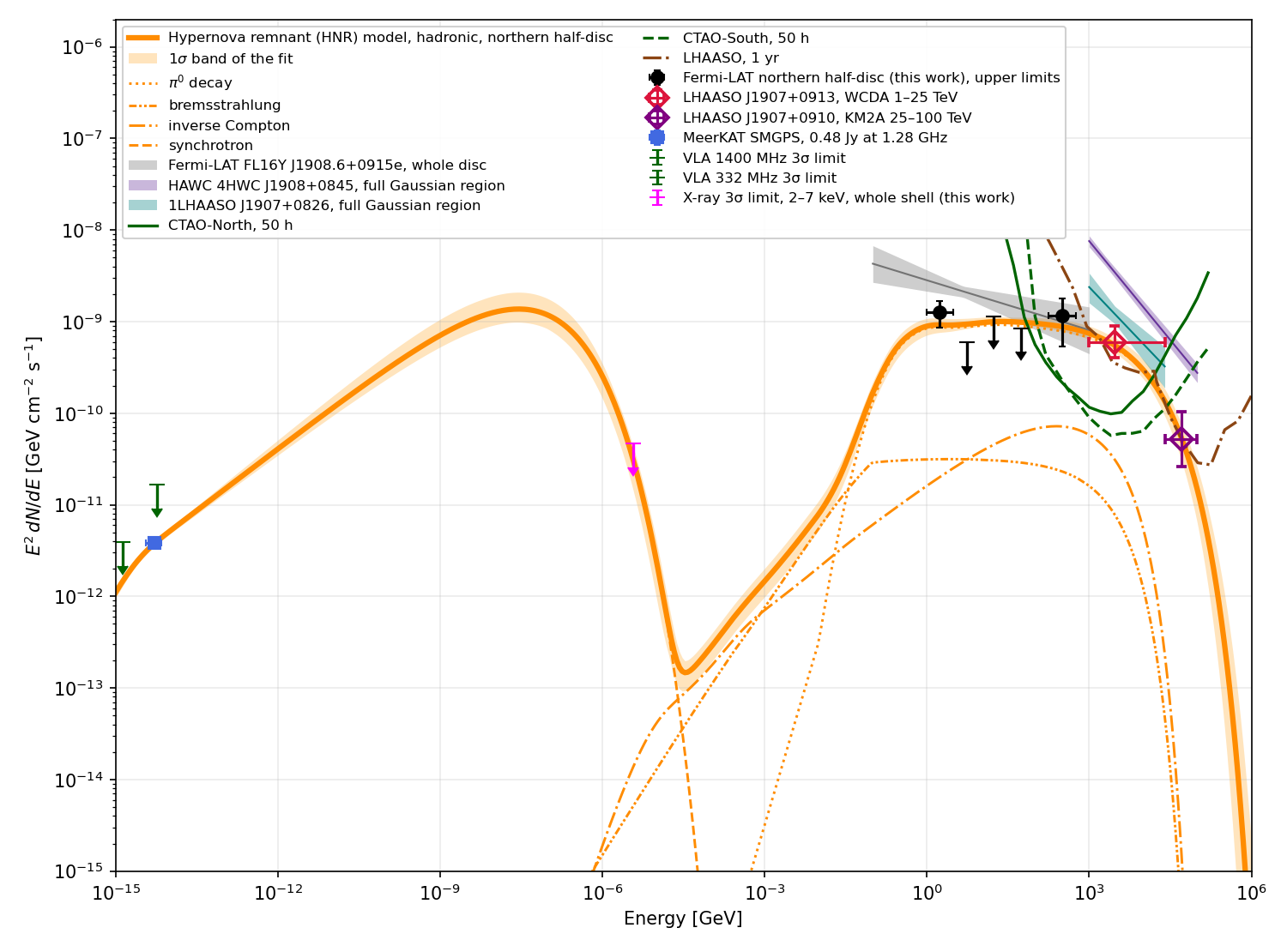}
 \caption{Broadband SED of the northern half-disc of \src{} in the HNR model (Table~\ref{tab:north-models}). Orange: best fit, total (solid) and its components: $\pi^0$ decay (dotted), bremsstrahlung (dashed with two dots), inverse Compton on the CMB, infrared and starlight photons (dash-dotted) and synchrotron (dashed); shading, $1\sigma$ band of the fit. Black circles: measured spectrum of the northern half-disc ($1\sigma$ errors, Table~\ref{tab:sed}, with horizontal bars showing the band width) and \lat{} $95$~per cent upper limits. Blue square: SMGPS flux density of G43.025+0.777 at $1.28$~GHz \citep{2025A&A...695A.144B}, with the $10$~per cent error used in the fit; the horizontal bar shows the MeerKAT L band ($0.86$--$1.71$~GHz). Open diamonds: our estimates for the LHAASO source inside the shell (Section~\ref{sec:lhaaso0913}): crimson, LHAASO~J1907+0913, WCDA, $3$~TeV ($1$--$25$~TeV); purple, its KM2A counterpart LHAASO~J1907+0910, $50$~TeV (above $25$~TeV); horizontal bars show the energy ranges, with the KM2A bar drawn up to $100$~TeV. Upper limits (bar and arrow): magenta, $2$--$7$~keV, unabsorbed, scaled from the rim to the whole shell (this work); green, VLA $3\sigma$ at $1.4$~GHz and $332$~MHz \citep{2002ApJ...566..378K}. Butterflies ($1\sigma$, errors on the normalisation and the index added in quadrature, Section~\ref{sec:lhaaso0913}): grey, \lat{}, whole disc; indigo, 4HWC~J1908+0845, and teal, 1LHAASO~J1907+0826, both for their full Gaussian regions. Dark green: CTAO differential sensitivity for $50$~h of observations of a point source (Prod5 instrument response functions; northern array at a zenith angle of $20^\circ$, solid; southern array at $40^\circ$, dashed) \citep{cta_prod5}; brown dash-dotted: one-year differential sensitivity of LHAASO to a Crab-like point source \citep{2019arXiv190502773B}. The sensitivity curves are drawn below $10^{-8}$~GeV~cm$^{-2}$~s$^{-1}$.}
 \label{fig:north_sed}
\end{figure}

\begin{figure}[H]
 \centering
 \includegraphics[width=\textwidth]{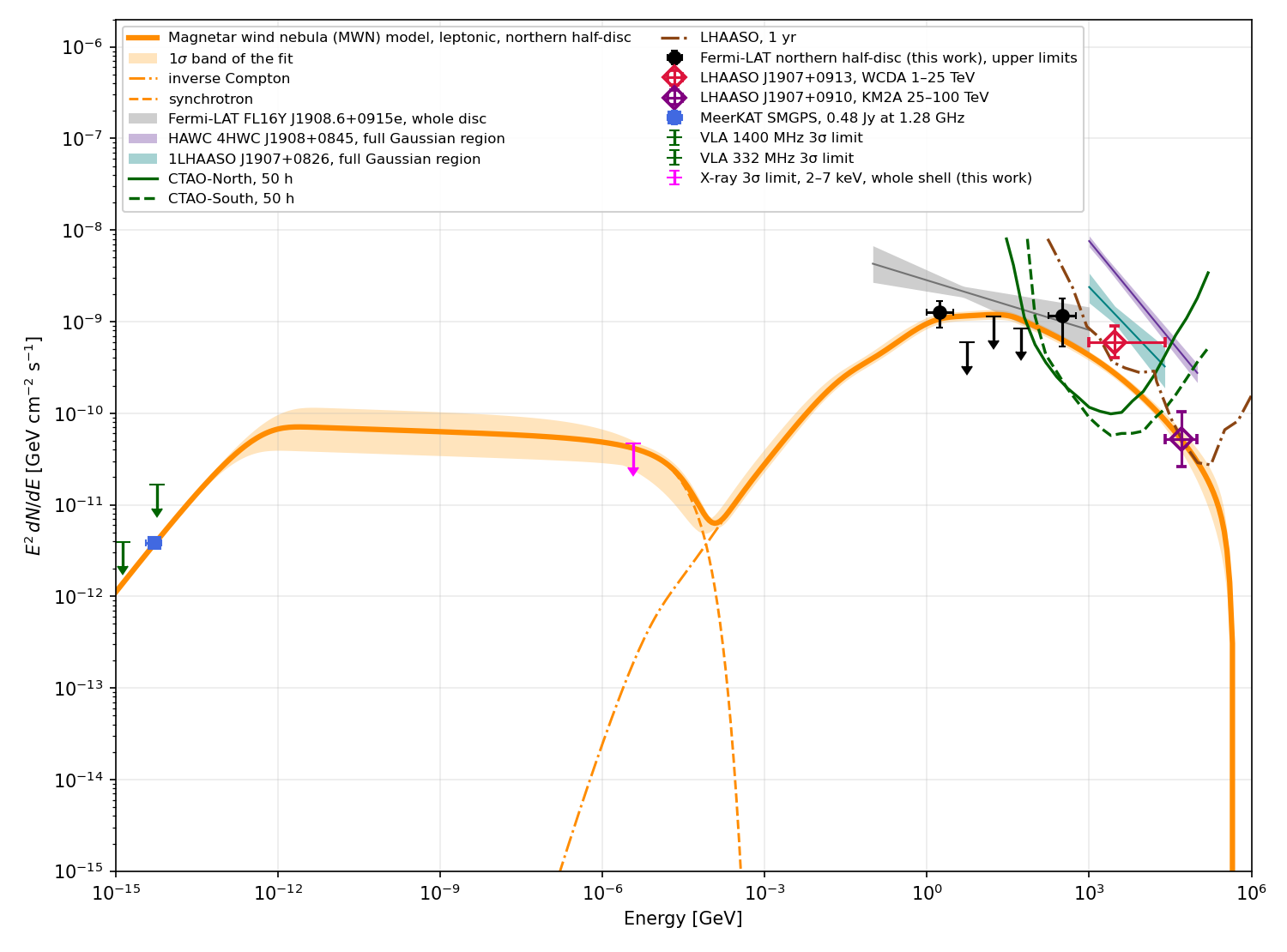}
 \caption{Same as Figure~\ref{fig:north_sed} for the MWN model: inverse Compton (dash-dotted) and synchrotron (dashed) emission of one electron population (Table~\ref{tab:north-models}).}
 \label{fig:north_sed_lepto}
\end{figure}

The radio and X-ray data are fitted together with the $\gamma$-ray data. The flux density of G43.025+0.777 in the SMGPS catalogue of extended sources is $F_\nu=0.477\pm0.013$~Jy at $\approx1.28$~GHz, after removal of $50$ compact sources inside the shell \citep{2025A&A...695A.144B}, as listed in Table~\ref{tab:co-radio-sources}. We use it as a detection with an error of $10$~per cent, which includes the calibration uncertainty and the extended flux that an interferometer may miss. No H\,{\sc ii} region of the WISE catalogue \citep{2014ApJS..212....1A} lies within the shell, and we assume that this flux is synchrotron emission of the shell. The X-ray limit for the whole shell is $7.5\times10^{-14}$~$\ergcms$, since the $7'$--$9.9'$ rim contains a fraction $\sqrt{1-(7/9.9)^2}=0.71$ of the flux of a thin shell. This limit, the \lat{} limits and the VLA $3\sigma$ limits of $6.2$ and $6.1$~mJy~arcmin$^{-2}$ at $1.4$~GHz and $332$~MHz \citep{2002ApJ...566..378K} multiplied by the area of the shell enter as one-sided constraints. They add to $\chi^2$ only when the model exceeds them. Each upper limit $F_{\rm UL}$ is then treated as a data point with an error of $F_{\rm UL}/2$, i.e. it adds $[(F_{\rm mod}-F_{\rm UL})/(F_{\rm UL}/2)]^2$ to $\chi^2$. All free parameters are fitted simultaneously by minimising $\chi^2=\sum_i[(\log F_{{\rm mod},i}-\log F_i)/\sigma_{\log F,i}]^2$, the sum over the data points of the squared differences between the logarithms of the model and the measured fluxes in units of their errors. The $1\sigma$ errors of the parameters are taken from the covariance matrix of the fit, i.e. in the approximation that the model is linear in its parameters near the minimum. For inverse Compton scattering we use the CMB, infrared emission with $T=107$~K and $1.19$~eV~cm$^{-3}$ and starlight with $T=7906$~K and $1.92$~eV~cm$^{-3}$, and we adopt $d=12.5$~kpc, all as in Paper 1. The $1\sigma$ bands in Figures~\ref{fig:north_sed} and \ref{fig:north_sed_lepto} are the envelope of all models with $\chi^2\le\chi^2_{\min}+1$ among parameter sets drawn from the fit covariance.

In the HNR model the protons have a power-law spectrum with an exponential cut-off (ECPL), ${\rm d}N/{\rm d}E=N_0(E/E_0)^{-\Gamma_{\mathrm{cr},p,1}}\exp(-E/E_{\rm cut})$, where $N_0$ is the normalisation at the reference energy $E_0=0.78$~TeV, and produce $\pi^0$ decay emission in the outer shell with $n_{\rm sh}=9.81$~cm$^{-3}$. In the same gas the electrons produce nonthermal bremsstrahlung, about $5$~per cent of the $\gamma$-ray flux. The electrons have the same index and a ratio $K_{ep}=0.0041$ to the protons at $E_0$, but their own cut-off $E_{{\rm cut},e}$, since radiative losses limit the electron energies much more than those of the protons. The fit is done in two stages. In the first stage, previous \textsc{naima} runs, including those of Paper 1, fixed the values of $n_{\rm sh}$, $K_{ep}$ and $E_0$. In the second stage these values are kept, and the remaining parameters are fitted to the present data. The free parameters are the normalisation, $\Gamma_{\mathrm{cr},p,1}$, $E_{\rm cut}$, $E_{{\rm cut},e}$ and $B$. The best-fit parameters are given in Table~\ref{tab:north-models}, and the model is shown in Figure~\ref{fig:north_sed}. The two \lat{} points and the two LHAASO estimates lie within $1\sigma$ of the model, which exceeds the $3$--$10$~GeV upper limit by a factor of $1.6$ and the $30$--$100$~GeV limit by $1.2$, and reaches $24$ and $54$~per cent of the VLA limits at $1.4$~GHz and $332$~MHz. The radio point fixes the magnetic field $B$. The model reaches the X-ray limit, which then fixes the electron cut-off, because the synchrotron emission of electrons above $\sim10$~TeV would exceed it. Behind a strong shock the field is compressed by a factor $\sqrt{11}$ (a strong shock compresses the gas by $r=4$, and an isotropic upstream field is compressed on average by $\sqrt{(1+2r^2)/3}=\sqrt{11}\approx3.3$), so the field in the bubble is $B_w\approx7\,\mu$G. $\Wp$ is about a quarter of the $5.1\times10^{50}$~erg found in Paper 1 for the whole region, since the northern half carries a third of the disc flux. The proton cut-off is set by the two LHAASO estimates alone; the other parameters do not depend on them (Table~\ref{tab:north-models}).

The MWN electrons have a spectrum similar to Paper 1, a broken power law with index $\Gamma_{\mathrm{cr},e,1}$ below the break energy $E_{\rm br}$, $\Gamma_{\mathrm{cr},e,2}$ above it, and an exponential cut-off (ECBPL). Two detected bands, three upper limits and two TeV estimates cannot constrain both indices and the break, so we fix $\Gamma_{\mathrm{cr},e,1}=1.49$ and $\Gamma_{\mathrm{cr},e,2}=3.04$ as in Paper 1 and fit the normalisation at $1$~TeV, $E_{\rm br}$, $E_{\rm cut}$ and $B$. The best-fit parameters are given in Table~\ref{tab:north-models}, and the model is shown in Figure~\ref{fig:north_sed_lepto}. $E_{\rm cut}$ is bounded only from below. The model exceeds the $3$--$10$~GeV upper limit by a factor of $2.0$ and the $30$--$100$~GeV limit by $1.2$, lies $2\sigma$ below the WCDA estimate and $10$~per cent above the X-ray limit. In the fitted field the synchrotron emission at $1.28$~GHz comes from electrons of $\approx6$~GeV, below the break, so the radio spectrum is flat. With the break fixed at $4.7$~GeV as in Paper 1, the radio emission would come from the steep part of the spectrum, the field would fall to $0.5\,\mu$G and the fit would be worse ($\chi^2=15.9$ against $9.3$ with one parameter fewer, $\Delta{\rm AIC}\approx+4.6$). $W_e$ is an order of magnitude below the $3.6\times10^{50}$~erg of the fit of Paper~1 to the whole region.

For any shape of the electron spectrum, the field of the MWN cannot be much higher than the fitted value: fields above $3$--$4\,\mu$G are disfavoured, as follows. The electrons of $100$--$200$~TeV that produce the $50$~TeV flux by inverse Compton scattering also radiate in the $2$--$7$~keV band, so the ratio of the X-ray limit to the TeV flux limits the ratio of the magnetic to the radiation energy density. With the indices fixed and $E_{\rm br}$ free, the best fit has $B=2.0\,\mu$G, and fields above $3$--$4\,\mu$G are disfavoured ($1.3\sigma$ at $3\,\mu$G and $2.2\sigma$ at $4\,\mu$G). A stronger field requires a lower electron cut-off to respect the X-ray limit, and the model then falls below the $50$~TeV flux.

 The hadronic model describes the data better: the leptonic model exceeds the $3$--$10$~GeV upper limit by a factor of $2$, and its $\chi^2$ is $9.3$ against $2.8$ with one free parameter fewer. The Akaike information criterion prefers the hadronic model by $\Delta{\rm AIC}\approx4.5$: the relative likelihood of the leptonic model, $\exp(-\Delta{\rm AIC}/2)\approx0.1$, so the data favour the hadronic model but do not exclude the leptonic one. The models differ by an order of magnitude in the field, and they predict different radio spectral indices $\alpha$, with $S_\nu\propto\nu^{-\alpha}$ (Table~\ref{tab:north-models}), typical of shell remnants for the HNR model and of pulsar wind nebulae for the MWN model. Thermal emission would give $\alpha\approx0.1$. The SMGPS flux and the VLA limit at $332$~MHz only require $\alpha\lesssim1.0$, so a measurement at a second frequency would distinguish the models.

\begin{table}[H]
\caption{NAIMA best-fit models of the SED of the northern half-disc of \src{}: a hypernova remnant (hadronic model, $i=p$) and a magnetar wind nebula (leptonic model, $i=e$), fitted to the measured spectrum of the half-disc (Table~\ref{tab:sed}).\label{tab:north-models}}
\footnotesize\setlength{\tabcolsep}{4pt}\renewcommand{\arraystretch}{1.12}\begin{tabularx}{\textwidth}{lCCCC}
\toprule {\bfseries\boldmath Parameter} & {\bfseries\boldmath HNR: ECPL ($i=p$)} & {\bfseries\boldmath HNR, LHAASO errors $\times3$} & {\bfseries\boldmath HNR, LHAASO as lower limits} & {\bfseries\boldmath MWN: ECBPL ($i=e$)} \\
\midrule
$E_{\mathrm{cr},i,\min}$ [GeV] (fixed) & 1 & 1 & 1 & 1 \\
$E_{\mathrm{cr},i,\max}$ [GeV] (fixed, \textsc{naima} default) & $10^7$ & $10^7$ & $10^7$ & $5.1\times10^5$ \\
$N_{0,i}$ [$10^{36}$~eV$^{-1}$] & \mbox{$9.7^{+2.2}_{-1.8}$} & \mbox{$9.2^{+2.8}_{-2.1}$} & \mbox{$8.9^{+7.2}_{-4.0}$} & \mbox{$16^{+14}_{-7}$} \\
$E_{\mathrm{cr},i,0}$ [TeV] (fixed)$^{c}$ & 0.78 & 0.78 & 0.78 & 1 \\
$E_{\mathrm{cr},i,\mathrm{br}}$ [TeV] & -- & -- & -- & \mbox{$0.070^{+0.046}_{-0.028}$} \\
$E_{\mathrm{cr},i,\mathrm{cut}}$ [TeV] & \mbox{$143^{+137}_{-70}$} & \mbox{$160^{+291}_{-103}$} & $>50$$^{e}$ & unconstrained \\
$E_{\mathrm{cr},e,\mathrm{cut}}$ [TeV] & \mbox{$5.8^{+2.3}_{-1.6}$} & \mbox{$6.2^{+3.0}_{-2.0}$} & \mbox{$6.4^{+6.4}_{-3.2}$} & -- \\
$\Gamma_{\mathrm{cr},i,1}$ & $2.10\pm0.09$ & $2.12\pm0.11$ & $2.13\pm0.22$ & 1.49 (fixed) \\
$\Gamma_{\mathrm{cr},i,2}$ & -- & -- & -- & 3.04 (fixed) \\
$W_p$ [$10^{50}$~erg] (calculated$^{a}$) & $1.2\pm0.3$ & $1.2\pm0.3$ & \mbox{$1.4^{+1.7}_{-0.2}$}$^{e}$ & -- \\
$W_e$ [$10^{47}$~erg] (calculated$^{a}$) & $4.2\pm1.1$ & $4.2\pm1.2$ & \mbox{$4.2^{+2.2}_{-0.7}$}$^{e}$ & $170\pm70$ \\
$n_\mathrm{sh}$ [cm$^{-3}$]$^{c}$ & 9.81 & 9.81 & 9.81 & -- \\
$K_{ep}$$^{c}$ & 0.0041 & 0.0041 & 0.0041 & -- \\
$B$ [$\mu$G] & \mbox{$23^{+8}_{-6}$} & \mbox{$22^{+9}_{-6}$} & \mbox{$22^{+13}_{-8}$} & \mbox{$2.0^{+1.2}_{-0.8}$} \\
Radio index $\alpha$ (0.33--1.28 GHz)$^{b}$ & 0.55 & 0.56 & 0.57 & 0.27 \\
$\chi^2$$^{d}$ & 2.8 & 2.7 & 2.6$^{e}$ & 9.3 \\
\bottomrule
\end{tabularx}
\noindent{\footnotesize\sloppy{Notes: The $\gamma$-ray data are the measured spectrum of the northern half-disc (Table~\ref{tab:sed}: two detected bands and three upper limits), the LHAASO estimates at 3 and 50~TeV, the MeerKAT flux density and the VLA and X-ray upper limits (Section~\ref{sec:xray:model}). Errors are $1\sigma$ from the covariance matrix of the fit. The HNR model exceeds the $3$--$10$~GeV upper limit by a factor of $1.6$ and the $30$--$100$~GeV limit by $1.2$; the MWN model by $2.0$ and $1.2$. In the third column the uncertainties of the two LHAASO estimates are enlarged to a factor of 3. In the fourth column the two estimates are replaced by lower limits on the flux, $1.1\times10^{-10}$ and $1.5\times10^{-11}$~GeV~cm$^{-2}$~s$^{-1}$ at 3 and 50~TeV (the estimates reduced for the uncertainty of the significance and the scatter of the calibration), which enter as one-sided constraints. $^{a}$Above $E_{\mathrm{cr},i,\min}$, for $d=12.5$~kpc. $^{b}$Of the model, $S_\nu\propto\nu^{-\alpha}$, computed with the electron spectrum extended down to $0.1$~GeV: at $B\approx20$~$\mu$G the emission at $0.33$~GHz comes from electrons of $\approx1$~GeV, so with the cut at $E_{\mathrm{cr},e,\min}=1$~GeV the index would be artificially flat ($0.30$). The fitted parameters do not depend on this choice, since below $1$~GHz the model is constrained only by the VLA upper limits and the flux at $1.28$~GHz changes by $2$~per cent. $^{c}$Fixed in the first stage of the fit (Paper~1). $^{d}$$\chi^2$ of the fit. The data are five measured points (two LAT bands, two LHAASO estimates, MeerKAT) and six upper limits (three LAT bands, X-ray, two VLA), which enter one-sidedly, with five (HNR) or four (MWN) free parameters; a formal number of degrees of freedom is therefore not defined, and the models are compared by $\Delta{\rm AIC}$. $^{e}$Three measured points (two LAT bands, MeerKAT), the two LHAASO lower limits and the six upper limits, with five free parameters. The proton cut-off is bounded only from below: $\chi^2=7.7$, 6.4, 3.6, 2.8 and 2.7 at $E_{\mathrm{cr},p,\mathrm{cut}}=10$, 30, 50, 70 and 100~TeV against the minimum 2.6, i.e. $E_{\mathrm{cr},p,\mathrm{cut}}>50$~TeV at $1\sigma$ and $>37$~TeV at 95~per cent; the upper errors of $W_p$ and $W_e$ (16th--84th percentiles) reflect the unconstrained cut-off.}}
\end{table}

\subsection {Modeling the W49A sky region
\label{sec:modeling-W49A-region}}
\label{sec:south}

 The spectrum of the southern half-disc is measured separately from that of the northern half (Table~\ref{tab:sed}).

The southern half disc contains the giant molecular cloud \wf{} with its star-forming complex. Its distance, $d=11.1$~kpc, comes from the trigonometric parallax of its water masers \citep{2013ApJ...775...79Z}. It is one of the most luminous star-forming regions in the Galaxy. It has $5$--$7\times10^{4}$~M$_\odot$ of young stars in several subclusters \citep{2005A&A...430..481H} inside a cloud of $1.1\times10^{6}$~M$_\odot$ within $60$~pc, with $2\times10^{5}$~M$_\odot$ in the central $6$~pc \citep{2013ApJ...779..121G}. Just outside the disc on the same side, $0.63^\circ$ from its centre, lies the supernova remnant W49B (G43.3$-$0.2, $d\approx11.3$~kpc~\citep{2022MNRAS.514..762H}), a separate \lat{} source (FL16Y~J1911.0+0905) and a TeV source \citep{2018A&A...612A...5H}. No TeV source lies inside the southern half. 1LHAASO~J1907+0826 (Section~\ref{sec:lhaaso0913}) is $0.92^\circ$ from the disc centre, outside the disc, and 4HWC~J1908+0845 (Section~\ref{sec:lhaaso0913}) partially covers both halves. At TeV energies the southern half has only upper limits.

The W49A molecular cloud cannot produce the observed emission if its $\gamma$-ray emission is produced only by the Galactic cosmic rays. For $2\times10^{5}$~M$_\odot$ (the cloud core) at $11.1$~kpc and the local cosmic-ray density, the expected photon flux above 1~GeV is $\approx$$2\times10^{-10}$~ph~cm$^{-2}$~s$^{-1}$, twelve times below the $2.4\times10^{-9}$~ph~cm$^{-2}$~s$^{-1}$ of the catalogue point source. A local accelerator is needed, and the obvious candidate is the young massive stellar cluster (YMSC) W49A. This is the situation of Westerlund~1 and of the Cygnus cocoon, where cosmic rays accelerated at the termination shock of the collective wind of the YMSC fill a region of tens of parsecs around the cluster \citep{2019NatAs...3..561A}.

\subsubsection {Modeling the YMSC W49A
\label{sec:modeling-W49A}}
\label{sec:south:models}

The catalogue point source FL16Y~J1910.2+0903 (4FGL~J1910.2+0904c in 4FGL-DR4) lies at $(l,b)=(43.129^\circ,-0.004^\circ)$, $1.9'$ ($6$~pc) from the radio centre of \wf{}, with a $95$~per cent error radius of $1.4'$, and $0.44^\circ$ from the disc centre. The catalogue associates it with WISE~G043.170$-$00.004, which is \wf{}, with $P_{\rm LR}=0.98$, and classifies it as a star-forming region. It is detected at $10.2\sigma$ with $\Gamma=2.304\pm0.074$ at a pivot energy of $7.2$~GeV and shows no significant curvature. Its $1$--$100$~GeV photon flux is $(2.42\pm0.36)\times10^{-9}$~ph~cm$^{-2}$~s$^{-1}$ (band spectrum in Table~\ref{tab:ps-sed}). 

 The similar YMSC Westerlund 1 has already partially dispersed the molecular cloud from which it formed, and its TeV $\gamma$-ray emission forms 
$R\sim 70$ pc ring around YMSC while GeV $\gamma$-ray emission covers an adjacent anisotropic (160 pc $\times$ 80 pc) stellar wind superbubble.

YMSC \wf{} is younger and still lies in the densest part of its cloud, so the cosmic rays interact with gas close to the cluster. The $\gamma$-ray emission of the point source is then produced by the cosmic rays in the cloud core, whereas that of extended source - by the cosmic ray- carrying wind and shock waves in the adjacent anisotropic superbubble around it.

We model both components with \textsc{naima} in the simplest way, as in Paper~1.

 The cloud core and the superbubble are fitted separately; only the proton cut-off of the superbubble is taken from the core (Table~\ref{tab:south-bubble}).

The protons have a power law spectrum with an exponential cut-off (ECPL) and produce $\gamma$-ray emission via $\pi^0$ decay. The electrons have the same spectrum, scaled by $K_{ep}=0.0041$, and 
produce bremsstrahlung in the same gas and inverse Compton emission on the CMB, infrared ($T=40$~K) and starlight photons. The infrared energy density is $40$~eV~cm$^{-3}$ in the cloud core and $2$~eV~cm$^{-3}$ in
the superbubble, and that of starlight is $2$~eV~cm$^{-3}$ in both. 
The mean molecular density within $6$~pc, 
$n_{\rm c}\approx4\times10^{3}$~cm$^{-3}$ \citep{2019A&A...622A..48R}, is used for the core, and the shell density $n_{\rm sh}=9.81$~cm$^{-3}$ for the superbubble. The values of $K_{ep}$ and $n_{\rm sh}$ are those fixed in the first stage of the two-stage fit of the northern model (Section~\ref{sec:xray:model}). The densities only set $W_p$ and $W_e$, which scale as $1/n$. For the synchrotron emission we assume a magnetic field of $70\,\mu$G in the core and $10\,\mu$G in the superbubble, close to the upper values of \citet{2010ApJ...725..466C} for these densities.

For the modelling of the point-like source FL16Y~J1910.2+0903 (the cloud core of \wf{}) we use its \lat{} spectrum from the catalogue (Table~\ref{tab:ps-sed}); its two upper limits below $300$~MeV and the H.E.S.S. limit discussed below enter as one-sided constraints.

An excess seen by H.E.S.S. towards \wf{} was not confirmed by the cross-check analysis \citep{2018A&A...612A...5H}, and the 95~per cent upper limit of the H.E.S.S. Galactic plane survey at its position \citep{2018A&A...612A...1H, 2018A&A...612A...5H} corresponds to $E^2\,{\rm d}N/{\rm d}E=1.43\times10^{-10}$ and $0.88\times10^{-10}$~$\gevcms$ at $1$ and $5$~TeV. These values correspond to the 95~per cent upper limit of the HGPS on the integral photon flux above 1~TeV within a correlation radius of $0.1^\circ$, $1.10\times10^{-13}$~cm$^{-2}$~s$^{-1}$ ($2.95\times10^{-13}$~cm$^{-2}$~s$^{-1}$ for $0.2^\circ$), converted to a differential flux for $\Gamma=2.3$ \citep{2018A&A...612A...1H}. The HGPS maps are not corrected for the containment of the point-spread function: a point source has only about 80~per cent of its flux within $0.1^\circ$ \citep{2018A&A...612A...1H}. The cloud core is point-like for H.E.S.S., so we apply the $0.1^\circ$ limit, divided by $0.8$, as a limit on the integral flux above 1~TeV, to the model of the point source; the superbubble contributes only about 5~per cent of the model flux inside the circle (for a uniform half-disc), which does not change the result. With this correction the integral flux above 1~TeV of the model with the proton cut-off at $E_{{\rm cut},p}=10$~TeV is about 10~per cent higher than the limit. This difference is smaller than the systematic uncertainty of the HGPS fluxes, about 30~per cent \citep{2018A&A...612A...1H}, so the model is consistent with the limit. Since this is a 95~per cent limit, the \lat{} data alone prefer slightly higher cut-offs, and the dependence on the spectral index assumed in the HGPS maps is not modelled, we take $E_{{\rm cut},p}\approx10$~TeV for the cloud core. The $0.2^\circ$ limit gives only $E_{{\rm cut},p}\lesssim$ $30$~TeV.

 Higher cut-offs give a flux above the H.E.S.S. limit. Lower values, $3$, $1$ and $0.3$~TeV, keep the model below
the limit but describe the \lat{} data worse and need harder indices, $2.2$, $2.0$ and $1.3$. We therefore fix $E_{\rm cut}$ at $10$~TeV.

\begin{figure}[H]
 \centering
 \includegraphics[width=\textwidth]{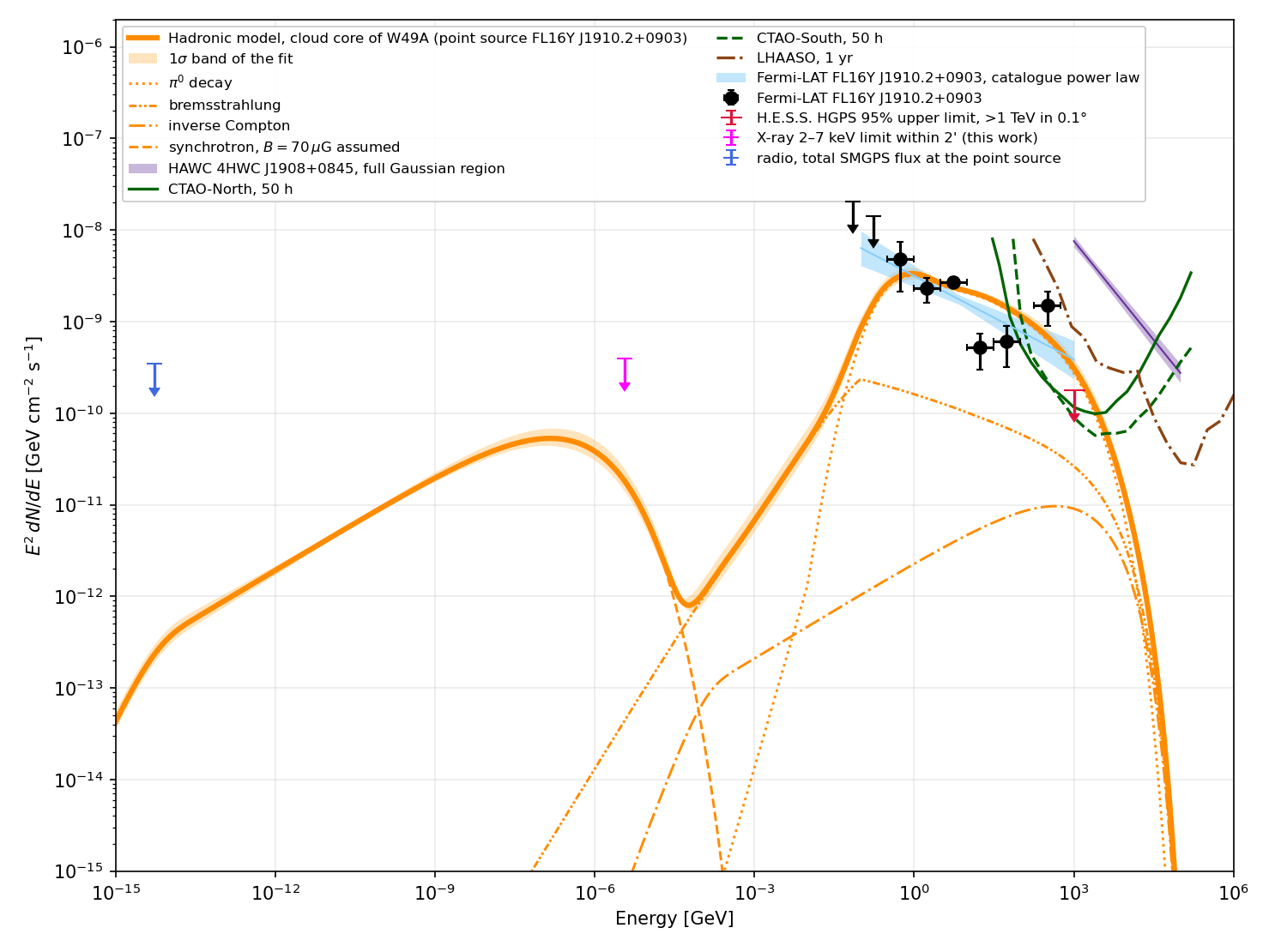}
 \caption{SED of the point source FL16Y~J1910.2+0903 (\wf{}) in the hadronic model fitted separately from the superbubble, and $E_{\rm cut}=10$~TeV (Table~\ref{tab:south-models}). Orange: best fit, total (solid) and its components: $\pi^0$ decay (dotted), bremsstrahlung (dashed with two dots), inverse Compton (dash-dotted) and synchrotron for an assumed $B=70\,\mu$G (dashed); shading, $1\sigma$ band of the fit. Black circles: catalogue bands of the point source ($1\sigma$ errors, with horizontal bars showing the band width) and \lat{} $95$~per cent upper limits. Upper limits (bar and arrow): crimson, H.E.S.S. Galactic plane survey at the position of \wf{}: the 95~per cent limit on the integral flux above 1~TeV within $0.1^\circ$, shown as the equivalent differential flux at 1~TeV for $\Gamma=2.3$, divided by $0.8$, the fraction of the flux of a point source within $0.1^\circ$ \citep{2018A&A...612A...1H, 2018A&A...612A...5H}; magenta, total \textit{Chandra} $2$--$7$~keV flux within $2'$ of \wf{}, mostly thermal; blue, the H\,{\sc ii} complex of \wf{} at $1.28$~GHz, thermal \citep{2025A&A...695A.144B}. Butterflies ($1\sigma$): light blue, catalogue power law of the point source; indigo, 4HWC~J1908+0845, full Gaussian region. Sensitivity curves as in Figure~\ref{fig:north_sed}.}
 \label{fig:south_pointsource_hadronic}
\end{figure}

The best-fit parameters are given in Table~\ref{tab:south-models}, and the model is shown in Figure~\ref{fig:south_pointsource_hadronic}.
 The proton index is $\Gamma_{\mathrm{cr},p,1}=2.30\pm0.07$. In the $1$--$100$~GeV band the model
has photon index of $2.23$ for the point source, in agreement with the catalogue value, $2.30\pm0.07$. 
The cut-off of the cloud core is not measured: the LAT data do not constrain it, and $E_{{\rm cut},p}\approx10$~TeV is the value at which the model is still consistent with the H.E.S.S. upper limit; a higher cut-off would give a flux above 1~TeV in excess of this limit. Under the current data and model assumptions, the \wf{} region does not show evidence for PeVatron activity. 
With the high cloud density of the core $n_{cl}=4\times 10^3$ cm$^{-3}$, the energy in protons of the point source is small,
$W_p=7\times10^{47}$~erg (observational flux is proportional to the product $W_p\times n_{cl}$). But if the filling factor of molecular gas $f_{c}<1$, total energy of cosmic rays increases to $W_p=7\times10^{49}f_{c,-2}$~erg.

 For the point source the $10$--$30$ and $30$--$100$~GeV bands lie $3.1\sigma$ and $1.8\sigma$ below the model, which gives the fit a probability of $7\times10^{-4}$ (Table~\ref{tab:south-models}), so the fit is poor ($\chi^2/{\rm ndf}=19.2/4$).

We should note some peculiarities of our fit. As follows from Tables~\ref{tab:sed} and~\ref{tab:ps-sed} and Figures~\ref{fig:south_pointsource_hadronic} and~\ref{fig:south_halfdisc}, both spectra of point-like and of extended sources in south half-disc 
demonstrate a dip in the 10--100~GeV region. The gamma-ray emission of the point-like source FL16Y J1910.2+0903 has already been studied in a number of works 
(in particular in \citet{Xin...Guo}), where it was noted that 
its spectrum corresponds to the hadronic mechanism, and the
dip of the spectrum at 10--100 GeV may be caused by a number of reasons. The 16-yr source list sets flag~10 for point-like source, which marks a poor agreement between its band fluxes and the catalogue spectral model \citep{2023arXiv230712546B}. The modelling of this energy range needs to take into account the contribution of diffuse emission and the nearby sources. In the neighbouring field of LHAASO~J1912+1014u, $1.4^\circ$ from \wf{}, the standard diffuse model leaves large residuals at $1.6$--$12.8$~GeV, and the catalogue sources flagged as coincident with interstellar clumps with soft spectra of unmodelled gas \citep{2026ApJ..1006...77M}. A separate analysis of the field would show whether the dip is real. Such an analysis is beyond the scope of this work.

\begin{table}[H]
\caption{NAIMA best-fit model of the SED of the point source FL16Y~J1910.2+0903, the cloud core of \wf{}.\label{tab:south-models}}
\small\begin{tabularx}{\textwidth}{lC}
\toprule
\textbf{Parameter} & \textbf{Cloud core of \wf{}: hadronic, ECPL} \\
\midrule
$E_{\mathrm{cr},p,\min}$ [GeV] (fixed) & 1 \\
$E_{\mathrm{cr},p,\max}$ [GeV] (fixed, \textsc{naima} default) & $10^7$ \\
$N_{0,p}$ [eV$^{-1}$] & $(3.1\pm0.5)\times10^{34}$ \\
$E_{\mathrm{cr},p,0}$ [TeV] (fixed)$^{b}$ & 0.78 \\
$E_{\mathrm{cr},p,\mathrm{cut}}$ [TeV] (fixed)$^{c}$ & 10 \\
$\Gamma_{\mathrm{cr},p,1}$ & $2.30\pm0.07$ \\
$W_p$ [erg] (calculated$^{a}$) & $7.0^{+1.3}_{-1.0}\times10^{47}$ \\
$W_e$ [erg] (calculated$^{a}$) & $2.9\times10^{45}$ \\
$n$ [cm$^{-3}$] (fixed)$^{d}$ & $4\times10^{3}$ \\
$K_{ep}$$^{b}$ & 0.0041 \\
Photon index, 1--100 GeV (model) & 2.23 \\
$\chi^2$/ndf & 19.2/4 \\
\bottomrule
\end{tabularx}
\noindent{\footnotesize{Notes: The data are the six detected catalogue bands of the point source, its two upper limits below 300~MeV and the H.E.S.S. HGPS $95$~per cent upper limit on the integral flux above 1~TeV within $0.1^\circ$, $1.10\times10^{-13}$~cm$^{-2}$~s$^{-1}$, divided by $0.8$ for the containment of the point-spread function (Section~\ref{sec:south:models}); the model exceeds this limit by 11~per cent. The probability of the fit is $7\times10^{-4}$, set by the dip of the point source at $10$--$100$~GeV (residuals $-3.1\sigma$ and $-1.8\sigma$). Electrons have the proton spectrum scaled by $K_{ep}$. $^{a}$Above $E_{\mathrm{cr},p,\min}$. $^{b}$Fixed in the first stage of the fit (Paper~1). $^{c}$ Set by the H.E.S.S. limit (Section~\ref{sec:south:models}). $^{d}$Mean density of the core, $2\times10^{5}$~M$_\odot$ within 6~pc.}}
\end{table}

\subsubsection {Modeling the W49A adjacent superbubble
\label{sec:modeling-supebubble}}

\begin{figure}[H]
 \centering
 \includegraphics[width=\textwidth]{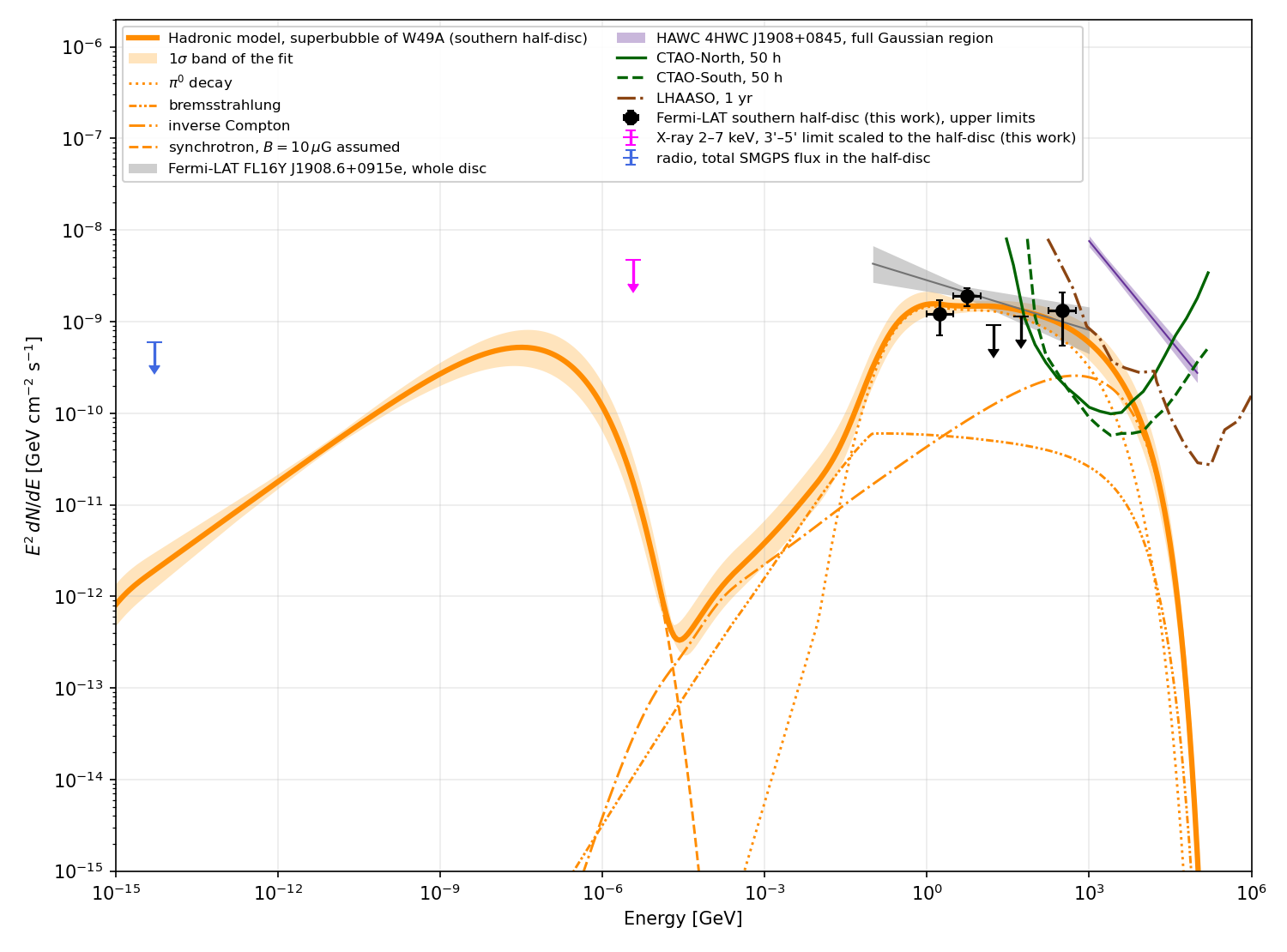}
 \caption{SED of the southern half-disc, the superbubble of \wf{}, in the hadronic model fitted separately, with the proton cut-off of the cloud core (Table~\ref{tab:south-bubble}), with synchrotron for an assumed $B=10\,\mu$G. Components as in Figure~\ref{fig:south_pointsource_hadronic}. Black circles: measured spectrum of the southern half-disc ($1\sigma$ errors, Table~\ref{tab:sed}) and \lat{} $95$~per cent upper limits. Upper limits: magenta, the \textit{Chandra} limit of the $3'$--$5'$ annulus around \wf{} scaled to the area of the half-disc; blue, total SMGPS flux in the half-disc at $1.28$~GHz, mostly from H\,{\sc ii} regions. Butterflies ($1\sigma$): grey, \lat{}, whole disc; indigo, 4HWC~J1908+0845, full Gaussian region. Sensitivity curves as in Figure~\ref{fig:north_sed}.}
 \label{fig:south_halfdisc}
\end{figure}

Modeling of extended superbubble adjacent to the W49A YMSC, presented in Table ~\ref{tab:south-bubble} and Figure~\ref{fig:south_halfdisc}, supports the hadronic mechanism of its $\gamma$-ray emission. 
Its energy in protons, $W_p=1.4\times10^{50}$~erg for $n_{\rm sh}=9.81$~cm$^{-3}$, is close to that of the northern half. The stellar winds of the YMSC can supply it. For $L_{\rm kin}\sim7\times10^{37}$~erg~s$^{-1}$ and an age of $2$~Myr their kinetic energy is $W_{\rm kin}\approx L_{\rm kin}t_{\rm age}\approx4.4\times10^{51}$~erg, and the protons of the superbubble take about $3$~per cent of it. The H.E.S.S. limit applies to the point source; the superbubble contributes only about 5~per cent of the flux inside the $0.1^\circ$ circle, and its cut-off is taken from the cloud core, on the assumption that the same cosmic rays of the YMSC fill both.

\begin{table}[H]
\caption{NAIMA best-fit model of the SED of the southern half-disc of \src{}, the superbubble of the YMSC \wf{}.\label{tab:south-bubble}}
\small\begin{tabularx}{\textwidth}{lC}
\toprule
\textbf{Parameter} & \textbf{Superbubble (southern half-disc): hadronic, ECPL} \\
\midrule
$E_{\mathrm{cr},p,\min}$ [GeV] (fixed) & 1 \\
$E_{\mathrm{cr},p,\max}$ [GeV] (fixed, \textsc{naima} default) & $10^7$ \\
$N_{0,p}$ [eV$^{-1}$] & $(1.1\pm0.3)\times10^{37}$ \\
$E_{\mathrm{cr},p,0}$ [TeV] (fixed)$^{b}$ & 0.78 \\
$E_{\mathrm{cr},p,\mathrm{cut}}$ [TeV] (fixed)$^{c}$ & 10 \\
$\Gamma_{\mathrm{cr},p,1}$ & $2.15\pm0.13$ \\
$W_p$ [erg] (calculated$^{a}$) & $(1.4\pm0.3)\times10^{50}$ \\
$W_e$ [erg] (calculated$^{a}$) & $5.6\times10^{47}$ \\
$n_\mathrm{sh}$ [cm$^{-3}$] (fixed)$^{d}$ & 9.81 \\
$K_{ep}$$^{b}$ & 0.0041 \\
Photon index, 1--100 GeV (model) & 2.05 \\
$\chi^2$/ndf & 3.5/1 \\
\bottomrule
\end{tabularx}
\noindent{\footnotesize{Notes: The data are the measured spectrum of the southern half-disc (Table~\ref{tab:sed}: three detected bands and two upper limits) and the 4HWC~J1908+0845 flux at 1 and 10~TeV as upper limits; the model reaches $0.05$--$0.08$ of the 4HWC flux. The probability of the fit is $0.06$. Electrons have the proton spectrum scaled by $K_{ep}$. $^{a}$Above $E_{\mathrm{cr},p,\min}$. $^{b}$Fixed in the first stage of the fit (Paper~1). $^{c}$Taken equal to the cut-off of the cloud core (Table~\ref{tab:south-models}), on the assumption that the same cosmic rays of the YMSC fill the core and the superbubble; with the cut-off free the fit gives $\Gamma_{\mathrm{cr},p,1}=2.23\pm0.21$ and does not constrain $E_{\mathrm{cr},p,\mathrm{cut}}$ from above. $^{d}$As in Paper~1.}}
\end{table}

\section{Discussion and Conclusions
\label{sec:discussion}}

In our work, we conducted an in-depth study of $\gamma$-ray sources in the SGR 1900+14--W49A sky region, begun in Paper 1. New multiwavelength observational data from recent years allow us to significantly improve our understanding of the sources and mechanisms that generate non-thermal radiation in this region. As mentioned above, radio observations have revealed several new SNR candidates, the center of one of which (G43.025+0.777) coincides with the position of the magnetar SGR1900+14 \citep{2025A&A...695A.144B}. In the multi-TeV band, the unidentified HAWC source 4HWC J1908+0845 \citep{2026arXiv260200263A} overlaps with the magnetar and its potential SNR G43.025+0.777, whereas a newly detected by LHAASO point-like source LHAASO J1907+0913 also coincides with the magnetar-connected SNR shell. Meantime, analysis of new Fermi-LAT data \citep{2026arXiv260222148B} concerning the considered in Paper 1 unidentified extended (with disc intensity profile) source 4FGL J1908.6+0915e / FL16Y J1908.6+0915e shows that the significance maps of the $R\approx0.6$~deg disc above 3~GeV have two maxima, near SGR~1900+14 and near W49A (Figure~\ref{fig:resid}), and that the disc is better described by two half-discs with different spectra (Section~\ref{sec:morphology}, Table~\ref{tab:sed}). These new results explain the extended source 4FGL J1908.6+0915e as a source confusion of two physical sources of comparable $\gamma$-ray spectral fluxes: magnetar SGR1900+14-connected SNR or MWN in the northern part and adjacent to the YMSC W49A stellar wind superbubble in the southern part of 4FGL J1908.6+0915e. Superbubble is created by the joint action of W49A stellar winds, collimated by large scale interstellar matter density gradient in the Galactic disc, analogously to the case of YMSC Westerlund 1 \citep{2025NatCo..16.10820L}. Moreover, similar to the Westerlund 1 case, HE $\gamma$-ray map of the W49A complex consists of a central maximum - point-like source FL16Y J1910.2+0903, connected with YMSC W49A, and extended stellar wind superbubble.

In both physical complexes: magnetar-connected SNR/MWN in the north and W49A in the south, there are favorable conditions for the operation of the cosmic PeVatrons. As analyzed in detail in Paper 1 in the north, magnetar-connected SNR/MWN at a considerable distance $d=12.5$ kpc should provide a Hypernova level $W\sim 10^{52}$ erg of SN explosion energy in order to explain observed $\gamma$-ray fluxes in the frame of hadronic or leptonic mechanisms. But, no significant VHE $\gamma$-ray emission from W49A was confirmed by H.E.S.S. \citep{2018A&A...612A...5H}, and, as shown in this paper, the W49A region is not bright in the multi-TeV band and, under the current data and model assumptions, does not show evidence for PeVatron activity.

Our main conclusions can be summarized as follows:

(i) in the SGR~1900+14--W49A sky region the unidentified extended source 4FGL J1908.6+0915e/ FL16Y J1908.6+0915e 
may be explained as a source confusion of two potential physical sources:
magnetar SGR1900+14-connected SNR or MWN and adjacent to the YMSC W49A stellar wind superbubble. YMSC W49A hosts point-like source 4FGL J1910.2+0904c/FL16Y J1910.2+0903.

(ii) In our two-component model, the new magnetar-connected extended source is responsible for the visible in radio SNR candidate G43.025+0.777, in the multi-TeV band contributes to the $\gamma$-ray flux of
HAWC extended source 4HWC J1908+0845 and sources the LHAASO J1907+0913. All these signatures of a PeVatron candidate require a total cosmic ray energy of order of $(1-5)\times 10^{50}$ erg, typical for Hypernova/MWN, supported by rotational energy of newborn millisecond magnetar. 

(iii) At the same time, both point-like 4FGL J1910.2+0904c/FL16Y J1910.2+0903 and new extended 
adjacent to the YMSC W49A sources are supported by a power supply from the kinetic luminosity of colliding 
stellar winds of hot stars in YMSC. Accelerated at multiple shock fronts, cosmic rays (protons and
electrons) inside YMSC are responsible for the 
4FGL J1910.2+0904c source. Cosmic rays in the 
extended region of anisotropically expanding in the circumcluster medium of decreasing density
(analogously to the Westerlund 1 case) are responsible for the new extended source. The absence of multi-TeV $\gamma$-ray emission from the W49A
region indicates the absence of conditions for accelerating cosmic rays to sub-PeV energies.

\vspace{6pt}

\authorcontributions{Conceptualization V.V., R.H., B.H.; methodology, V.V., L.Z., O.G., V.B. and B.H.; software, L.Z. and V.V.; formal analysis, V.V., L.Z., R.H. and B.H.; data curation, L.Z., V.V.; investigation, V.V., L.Z., B.H., O.G., V.B., R.H. and A.S.; validation, R.H. and A.S.; visualization, L.Z., V.V., A.S.; writing---original draft preparation, B.H., V.V., and L.Z.; writing---review and editing, V.V., B.H., L.Z., O.G., V.B., R.H. and A.S.; supervision, B.H., L.Z., V.V. All authors have reviewed, discussed, and~commented on the present results. All authors have read and agreed to the published version of the~manuscript.}

\funding{
The work of B.H. was supported by The National Research Foundation of Ukraine (project 2023.03/0149), Ukraine.
The work of R.H. was supported by the Ministry of Education and Science of Ukraine (project 0124U001660). 
The work of O.G. was carried out within the framework of the research project State Registration No. 0125U002256.
The work of L.Z. has received funding from the MSCA4Ukraine project, grant number Ref 1.4-UKR-1245772-MSCA4Ukraine, which is funded by the European Union. The views and opinions expressed are, however, those of the author(s) only and do not necessarily reflect those of the European Union, the European Research Executive Agency, or the MSCA4Ukraine Consortium. Neither the European Union nor the European Research Executive Agency, nor the MSCA4Ukraine Consortium as a whole, nor any individual member institutions of the MSCA4Ukraine Consortium, can be held responsible for them. 
The work of L.Z. was partially supported by the Ministry of Education and Science of Ukraine under the project ``Search for dark matter and particles beyond the Standard Model'' (0125U002260). 
This research was conducted with support from the Centre for the Collective Use of Scientific Equipment "Laboratory of High Energy Physics and Astrophysics" of Taras Shevchenko National University of Kyiv.}

\dataavailability{The X-ray data analysed in this work are publicly available. The \textit{XMM-Newton} observations of \magnetar{} (ObsIDs 0305580101, 0305580201, 0410580101, 0410580201, 0506430101, 0506430201 and 0790610101) and of W49B (ObsIDs 0084100401, 0084100501, 0084100601, 0724270101 and 0724270201) can be retrieved from the \textit{XMM-Newton} Science Archive (\url{https://nxsa.esac.esa.int}), and the \textit{Chandra} observations of \magnetar{} (ObsIDs 1954 and 6731) and of \wf{} (ObsIDs 5893 and 6355) from the \textit{Chandra} Data Archive (\url{https://cxc.harvard.edu/cda}). The radio flux densities were taken from the SMGPS extended-source catalogue \citep{2025A&A...695A.144B}, available at the CDS (\url{https://cdsarc.cds.unistra.fr/viz-bin/cat/J/A+A/695/A144}). The LHAASO flux estimates were derived from the published detection significances and the 1LHAASO catalogue \citep{2024ApJS..271...25C, 2024arXiv241008988C, 2026arXiv260625054C}. The spectral energy distributions were modelled with the public code \textsc{naima} \citep{2015ICRC...34..922Z}. The processed X-ray images and count limits, the fitted model parameters and the modelling scripts are available from the corresponding author on reasonable request.
}

\acknowledgments{We are deeply grateful to the four anonymous referees for the critical assessment of our analysis, the attention to the physics of the problem, the thorough examination of the methodology and the interpretation, and the careful reading of the text. Their constructive comments led us to revise the analysis and improved the paper substantially.}

\conflictsofinterest{The authors declare no conflicts of interest.}

\appendixtitles{yes}
\appendixstart
\appendix
\section[\appendixname~\thesection]{Counts and flux maps of the region}\label{app:maps}

\begin{figure}[H]
\includegraphics[width=\textwidth]{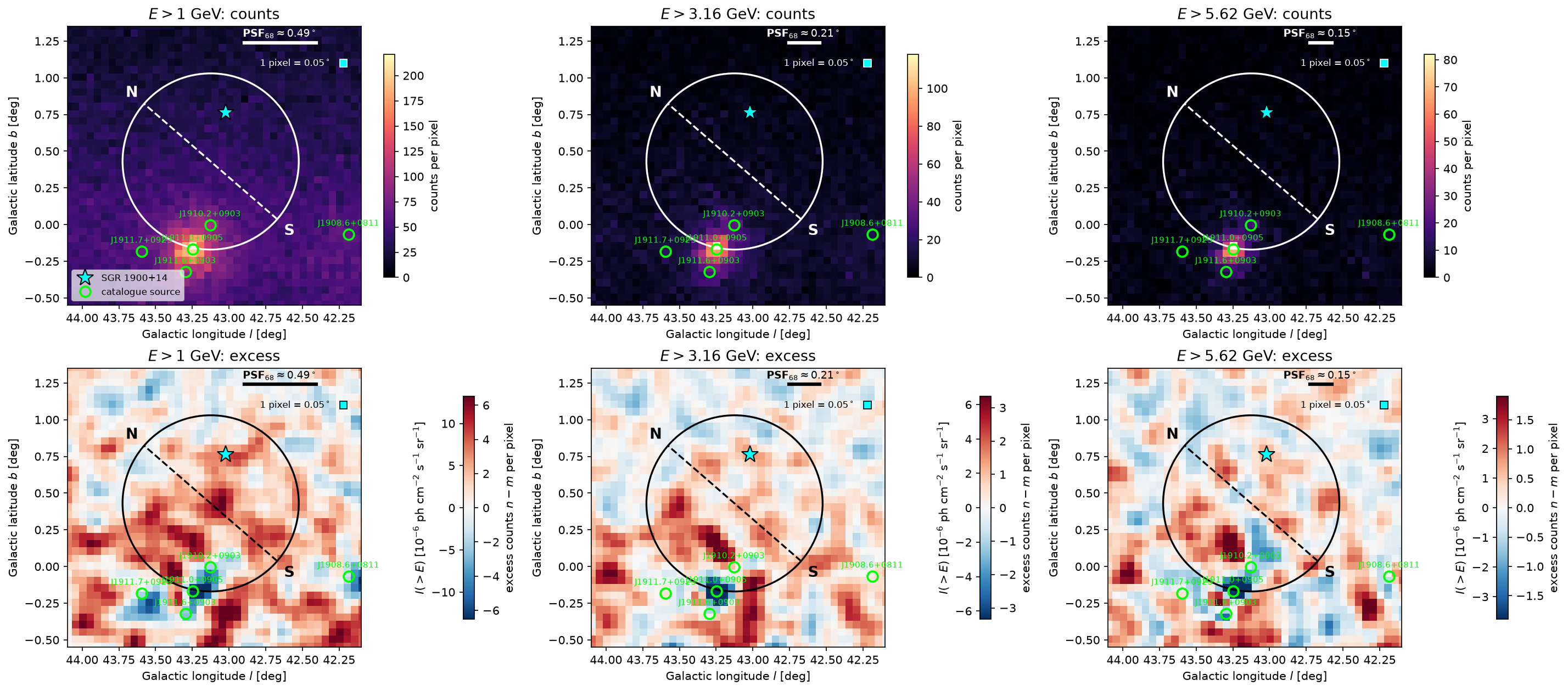}
\caption{Counts and residual maps of the region of Figure~\ref{fig:resid}, for photons above $1$, $3.16$ and $5.62$~GeV (columns; the thresholds of Figure~\ref{fig:resid}, at the bin edges of the analysis). Top: counts $n$ per $0.05^\circ$ pixel (source and background), all four PSF event types summed, from the 16~yr data set (Section~\ref{sec:morphology}). Bottom: residual counts $n-m$ per pixel (left-hand colour scale), where $m$ is the number of counts predicted by the best-fit model without \src{}, the background model of Figure~\ref{fig:resid}, and the corresponding intensity $I(>E)=(n-m)/(\varepsilon\,\Omega_\mathrm{pix})$ (right-hand scale), where $\varepsilon\simeq6.6\times10^{11}$~cm$^2$~s is the exposure for the spectrum of the disc, uniform over the region to $1\%$, and $\Omega_\mathrm{pix}=7.6\times10^{-7}$~sr is the solid angle of a pixel. The residual maps are smoothed with a Gaussian of $\sigma=1$~pixel for display. Circle and dashed line: \src{} and its division; cyan star: \magnetar{}; green circles: the catalogue sources in the field, among them FL16Y~J1910.2+0903 (\wf{}) and FL16Y~J1911.0+0905 (W49B). The bar gives the 68~per~cent containment radius of the PSF above each threshold and the cyan square one pixel.\label{fig:maps}}
\end{figure}

Figure~\ref{fig:maps} complements the significance maps of Figure~\ref{fig:resid} with the counts and with the residual counts and intensity in the same region and energy ranges. The counts maps are dominated by the supernova remnant W49B, just outside the disc, which the background model reproduces. The residual maps show, pixel by pixel, the photons that the best-fit model without \src{} does not account for, i.e. the emission whose significance is mapped in Figure~\ref{fig:resid}.

\begin{adjustwidth}{-\extralength}{0cm}
\reftitle{References}
\bibliography{references}
\end{adjustwidth}

\PublishersNote{}
\end{document}